\documentclass[conference]{./templates/ndss/IEEEtran}
\usepackage{etoolbox}

\usepackage{hyphenat}

\usepackage{balance}

\usepackage{url}
\usepackage{cite}
\usepackage{multirow}
\usepackage{subfigure}
\usepackage{verbatim}
\usepackage{algorithm}
\usepackage{algorithmic}
\usepackage{listings}
\usepackage{color}
\usepackage{tabu}
\usepackage{booktabs}
\usepackage{flushend}

\usepackage{footmisc}
\usepackage{comment}

\usepackage{graphicx}

\usepackage{float}
\usepackage{framed}

\usepackage{calc}
\usepackage{ifthen}
\usepackage{tikz}
\usepackage{datapie}
\usepackage{fp}
\usepackage{pifont}

\usepackage{caption}

\def\myhyperrefcolor{black}

\usepackage[
colorlinks=true,
citecolor=\myhyperrefcolor,
filecolor=\myhyperrefcolor,
linkcolor=\myhyperrefcolor,
urlcolor=\myhyperrefcolor,
pagecolor=\myhyperrefcolor,
linkbordercolor={1 1 1},
bookmarksopen=true,
bookmarksnumbered=true,
frenchlinks=false,
bookmarks=true
]{hyperref}

\newcounter{t0d0_counter}
\newcommand{\notodo}[1]{
}

\newcounter{pr00f_counter}

\newcommand\printpercent[2]{\the\numexpr#1*100/#2\%}

  \renewenvironment{thebibliography}[1]{%
    \begin{oldthebibliography}{#1}%
      \setlength{\parskip}{0ex}%
      \setlength{\itemsep}{0ex}%
  }%
  {%
    \end{oldthebibliography}%
  }

\renewcommand{\figurename}{Figure}

\usepackage{chngcntr}
\counterwithout{paragraph}{subsubsection} 
\counterwithin{paragraph}{subsection} 

\newcommand{\countfirmwarewebTP}{1925}
\newcommand{\CountFirmwareWebTPVendors}{54}

\newcommand{\countfirmwareforemulation}{1580}
\newcommand{\countrootfsforemulationBeforeFixingFS}{1754}
\newcommand{\countrootfsforemulation}{1982}
\newcommand{\countrootfsforemulationFixedFS}{228} 
\newcommand{\CountFirmwareEmulatedOK}{488}
\newcommand{\CountFirmwareEmulatedNOTOK}{1092}
\newcommand{\CountVendorsEmulatedOK}{17}
\newcommand{\CountFirmwareEmulatedWebServerStarted}{246}
\newcommand{\CountFirmwareEmulatedWebServerNOTStarted}{242}
\newcommand{\CountVendorsEmulatedWebServerStarted}{11}
\newcommand{\CountFirmwareHostedDynArachni}{515}

\newcommand{\CountFirmwareForEmulationVendors}{49}

\newcommand{\PrcntFirmwareBinaryCGI}{57\%}

\newcommand{\CountVulnsVendorsTotal}{13}
\newcommand{\countfirmwarestotal}{185}

\newcommand{\CountFirmwareDynVulnTotal}{58} 

\newcommand{\CountFirmwareDynVulnTotalHigh}{45} 
\newcommand{\CountFirmwareDynVulnTotalLow}{23} 
\newcommand{\CountFirmwareDynVulnCmdInj}{21}
\newcommand{\CountFirmwareDynVulnXSS}{32}
\newcommand{\CountFirmwareDynVulnCSRF}{37}
\newcommand{\CountFirmwareDynVulnBackup}{1}
\newcommand{\countfirmwaredynvulncookiehttp}{9}
\newcommand{\countfirmwaredynvulnxcontenttype}{23}
\newcommand{\countfirmwaredynvulnxframeoptions}{23}
\newcommand{\CountFirmwareDynVulnErrorDiscl}{1}

\newcommand{\CountFirmwareDynVulnTotalHighHosted}{307}

\newcommand{\CountFirmwareDynVulnCmdInjHosted}{15}

\newcommand{\CountFirmwareDynVulnXSSHosted}{13}

\newcommand{\CountFirmwareDynVulnCSRFHosted}{307}

\newcommand{\CountDynVulnsTotal}{6068} 
\newcommand{\CountDynVulnsTotalHigh}{225}
\newcommand{\CountDynVulnsTotalLow}{5843} 
\newcommand{\CountDynVulnsCmdInj}{51} 
\newcommand{\CountDynVulnsXSS}{90}
\newcommand{\CountDynVulnsCSRF}{84}
\newcommand{\CountDynVulnsBackup}{2}
\newcommand{\countdynvulncookiehttp}{9}
\newcommand{\countdynvulnxcontenttype}{2938}
\newcommand{\countdynvulnxframeoptions}{2893}
\newcommand{\CountDynVulnAppErrorDiscl}{1}

\newcommand{\CountFirmwarePhp}{150}
\newcommand{\countfirmwarephpfwripsvuln}{145}
\newcommand{\countfirmwarephpripstotal}{9046}

\newcommand{\CountVulnsConclusion}{9271} 

\newcommand{\countfirmwarephpripscodeexecution}{141}
\newcommand{\countfirmwarephpripshttpresponsesplitting}{127}
\newcommand{\countfirmwarephpripscommandexecution}{938}
\newcommand{\countfirmwarephpripsheaderinjection}{1}
\newcommand{\countfirmwarephpripspossibleflowcontrol}{171}
\newcommand{\countfirmwarephpripsunserialize}{119}
\newcommand{\countfirmwarephpripssqlinjection}{442}
\newcommand{\countfirmwarephpripspopgadgets}{4}
\newcommand{\countfirmwarephpripsfiledisclosure}{461}
\newcommand{\countfirmwarephpripsfilemanipulation}{1129}
\newcommand{\countfirmwarephpripsfileinclusion}{513}
\newcommand{\countfirmwarephpripscrosssitescripting}{5000}

\newcommand{\countfirmwarefilesphpripscodeexecution}{21}
\newcommand{\countfirmwarefilesphpripshttpresponsesplitting}{27}
\newcommand{\countfirmwarefilesphpripscommandexecution}{41}
\newcommand{\countfirmwarefilesphpripsheaderinjection}{1}
\newcommand{\countfirmwarefilesphpripspossibleflowcontrol}{56}
\newcommand{\countfirmwarefilesphpripsunserialize}{15}
\newcommand{\countfirmwarefilesphpripssqlinjection}{10}
\newcommand{\countfirmwarefilesphpripspopgadgets}{4}
\newcommand{\countfirmwarefilesphpripsfiledisclosure}{87}
\newcommand{\countfirmwarefilesphpripsfilemanipulation}{98}
\newcommand{\countfirmwarefilesphpripsfileinclusion}{40}
\newcommand{\countfirmwarefilesphpripscrosssitescripting}{143}

\newcommand{\countfirmwaremanualfwtotal}{19}
\newcommand{\countfirmwaremanualfwprivesc}{19}
\newcommand{\countfirmwaremanualfwunauthconfdown}{19}
\newcommand{\countfirmwaremanualfwunencryptconf}{19}

\newcommand{\CountFirmwaresOtherServices}{207}

\newcommand{\countfirmwareswithHTTPScerts}{363}

\newcommand{\CountFirmwareEmulatedWebserverStartedHTTPS}{60}

\newcommand{\CountFirmwareEmulatedOKByFixingFS}{9}
\newcommand{\CountFirmwareWebServerOKByFixingFS}{26}

\newcommand{\CountFWArchArmel}{666}
\newcommand{\CountFWArchMips}{374}
\newcommand{\CountFWArchMipsel}{323}
\newcommand{\CountFWArchBflt}{98}
\newcommand{\CountFWArchCris}{315}
\newcommand{\CountFWArchPowerpc}{60}
\newcommand{\CountFWArchIntel}{32}

\newcommand{\CountFWArchDLinkDirElfMSB}{21}
\newcommand{\CountFWArchUnknown}{22}

\newcommand{\CountFWArchArmelEmulationOk}{258}
\newcommand{\CountFWArchMipsEmulationOk}{104}
\newcommand{\CountFWArchMipselEmulationOk}{126}

\newcommand{\CountFWArchArmelWebserverOk}{136}
\newcommand{\CountFWArchMipsWebserverOk}{42}
\newcommand{\CountFWArchMipselWebserverOk}{68}

\newcommand{\CountFirmwareWebTechHtml}{242}
\newcommand{\CountFirmwareWebTechCgi}{140}
\newcommand{\CountFirmwareWebTechPhp}{5}
\newcommand{\CountFirmwareWebTechPerl}{8}

\newcommand{\CountFirmwareWebTechShell}{26} 

\newcommand{\CountFirmwareEmulatedWebServerStartedMinihttpd}{90}
\newcommand{\CountFirmwareEmulatedWebServerStartedLighttpd}{75}
\newcommand{\CountFirmwareEmulatedWebServerStartedBoa}{10}
\newcommand{\CountFirmwareEmulatedWebServerStartedThttpd}{8}

\newcommand{\EmulatedNOTOKRandSampl}{88} 

\newcommand{\EmulatedNOTOKRandSamplChrootDetectFail}{52}
\newcommand{\EmulatedNOTOKRandSamplChrootDetectFailPct}{59.1\%}
\newcommand{\EmulatedNOTOKRandSamplChrootDetectFailConfInt}{9.8\%}

\newcommand{\EmulatedNOTOKRandSamplChrootFail}{36}
\newcommand{\EmulatedNOTOKRandSamplChrootFailPct}{40.9\%}
\newcommand{\EmulatedNOTOKRandSamplChrootFailConfInt}{9.8\%}

\newcommand{\EmulatedNOTOKRandSamplChrootFailExecFormat}{10}
\newcommand{\EmulatedNOTOKRandSamplChrootFailExecFormatPct}{11.3\%}
\newcommand{\EmulatedNOTOKRandSamplChrootFailExecFormatConfInt}{6.3\%}

\newcommand{\EmulatedNOTOKRandSamplChrootFailFS}{26}
\newcommand{\EmulatedNOTOKRandSamplChrootFailFSPct}{29.5\%}
\newcommand{\EmulatedNOTOKRandSamplChrootFailFSConfInt}{9.1\%}

\newcommand{\EmulatedNOTOKRandSamplFixEASY}{62}
\newcommand{\EmulatedNOTOKRandSamplFixEASYPct}{70.4\%}
\newcommand{\EmulatedNOTOKRandSamplFixEASYConfInt}{9.1\%}
\newcommand{\EmulatedNOTOKRandSamplFixEASYPctMin}{61.3\%}

\newcommand{\WebServerNOTOKRandSampl}{69} 

\newcommand{\WebServerNOTOKRandSamplDeviceFail}{45}
\newcommand{\WebServerNOTOKRandSamplDeviceFailPct}{65.2\%}
\newcommand{\WebServerNOTOKRandSamplDeviceFailConfInt}{9.5\%}

\newcommand{\WebServerNOTOKRandSamplInitFail}{15}
\newcommand{\WebServerNOTOKRandSamplInitFailPct}{21.8\%}
\newcommand{\WebServerNOTOKRandSamplInitFailConfInt}{8.2\%}

\newcommand{\WebServerNOTOKRandSamplWebservFail}{9}
\newcommand{\WebServerNOTOKRandSamplWebservFailPct}{13.0\%}
\newcommand{\WebServerNOTOKRandSamplWebservFailConfInt}{6.7\%}

\newcommand{\WebServerNOTOKRandSamplFixEASY}{24}
\newcommand{\WebServerNOTOKRandSamplFixEASYPct}{34.8\%}
\newcommand{\WebServerNOTOKRandSamplFixEASYConfInt}{9.6\%}
\newcommand{\WebServerNOTOKRandSamplFixEASYPctMin}{25.2\%}

\begin{document}

\title{Automated Dynamic Firmware Analysis at Scale:\\ A Case Study on Embedded Web Interfaces}

\author{

\IEEEauthorblockN{Andrei Costin}
\IEEEauthorblockA{EURECOM\\
Sophia Antipolis, France\\
\href{mailto:costin@eurecom.fr}{costin@eurecom.fr}
}
\and
\IEEEauthorblockN{Apostolis Zarras}
\IEEEauthorblockA{Ruhr-University Bochum\\
Germany\\
\href{mailto:apostolis.zarras@rub.de}{apostolis.zarras@rub.de}
}
\and
\IEEEauthorblockN{Aur\'elien Francillon}
\IEEEauthorblockA{EURECOM\\
Sophia Antipolis, France\\
\href{mailto:francill@eurecom.fr}{francill@eurecom.fr}
}

}

\IEEEoverridecommandlockouts
\makeatletter\def\@IEEEpubidpullup{9\baselineskip}\makeatother
\IEEEpubid{\parbox{\columnwidth}{
}
\hspace{\columnsep}\makebox[\columnwidth]{}}

\maketitle

\begin{abstract}  
Embedded devices are becoming more widespread, interconnected, and
web-enabled than ever. However, recent studies showed that these
devices are far from being secure. Moreover, many embedded systems
rely on web interfaces for user interaction or administration. 
Unfortunately, web security is known to be difficult, and therefore the web interfaces of
embedded systems represent a considerable attack surface.

In this paper, we present the \emph{first fully automated framework} that 
applies dynamic firmware analysis techniques to achieve, in a scalable manner, 
automated vulnerability discovery within embedded firmware images. 
We apply our framework to study the security of embedded web
interfaces running in Commercial Off-The-Shelf (COTS) embedded
devices, such as routers, DSL/cable modems, VoIP phones, IP/CCTV
cameras.  We introduce a methodology and implement a
scalable framework for discovery of vulnerabilities in embedded web
interfaces regardless of the vendor, device, or architecture.
To achieve this goal, our framework performs full system emulation to achieve 
the execution of firmware images in a software-only environment, i.e., without 
involving any physical embedded devices. Then, we analyze 
the web interfaces within the firmware using both static and dynamic tools. 
We also present some interesting case-studies, and discuss the main challenges 
associated with the dynamic analysis of firmware images and their web 
interfaces and network services. 
The observations we make in this paper shed light on an important
aspect of embedded devices which was not previously studied at a large
scale. Insights from this paper can help users, programmers and
auditors in efficiently testing and securing their Internet enabled
embedded devices.

We validate our framework by testing it on \countfirmwarewebTP{} 
firmware images from \CountFirmwareWebTPVendors{} different vendors. 
We discover important vulnerabilities in \countfirmwarestotal{} firmware 
images, affecting nearly a quarter of vendors in our dataset. 
We also perform comprehensive failure analysis. We show that 
by applying \emph{relatively easy} fixes during corrective maintenance
it is possible to remediate at least \EmulatedNOTOKRandSamplFixEASYPctMin{} 
of emulation failures and at least \WebServerNOTOKRandSamplFixEASYPctMin{} of 
web interface launch failures. 
These experimental results demonstrate the effectiveness of our approach.

\end{abstract}

\section{Introduction}
\label{sec:introduction}

Embedded devices are present in many complex systems, like cars,
planes, and programmable logic controllers. Such devices also appear
massively in customer products such as network gateways and IP cameras.
Those devices are becoming more pervasive and ``invade'' our 
lives under many different forms (e.g., home automation, smart
TVs).

Embedded systems, in particular Small Office/Home Office (SOHO)
devices, are often known to be insecure~\cite{geer_bh,
ieee_sp_state_of_embedded}. Their lack of security may be the
consequence of the harsh market competition. For instance,
the time to market is crucial and the competition puts high pressure
on the design and production costs, and enforces short release 
timelines. Vendors try to provide as many 
features as possible to differentiate products, while customers
do not necessarily look for the most secure products. 

Some embedded systems have clear and well-defined security goals, such as 
the pay-TV smart cards and the Hardware Security Modules (HSM). 
Therefore, such devices are rather secure. However, 
many embedded systems are not designed with a clear threat model in mind.
This gives little motivation to manufacturers to invest time and money 
in securing them. This fact motivated several researchers
to evaluate the state of security of such embedded
devices~\cite{costin2014large, cui-acsac2011-PreyToHunter, 
internetcensus2012, Tripwire, elie-bh2009-EmbedInterfMassInsec, DBLP:journals/corr/NiemietzS15}.

Moreover, during the past few years, embedded devices became more connected
forming what is called the Internet of Things (IoT). Such devices are often
put online by composition; attaching a communication interface to an
existing (insecure) device.
Most of these devices lack the user interface of desktop computers
(e.g., keyboard, video, mouse), but nevertheless need to be 
configured and maintained. Albeit some devices rely on custom protocols used by ``thick''
clients or even legacy interfaces (i.e., telnet), the web quickly
became the universal administration interface.
Thus, the firmware of these devices often embed a web server running 
web applications, for the rest of this paper, we will refer to these as 
\emph{embedded web interfaces}.

It is well known that making secure web applications  is not a trivial task.
In particular, researchers showed that more than 70\% of vulnerabilities 
are hosted in the (web) application layer~\cite{gartner70percent}. 
Attackers who are familiar with this fact 
use various techniques to exploit web applications.
Well known vulnerabilities, such as SQL injection~\cite{boyd2004sqlrand}
or Cross Site Scripting (XSS)~\cite{vogt2007cross}, 
constitute a significant portion of the vulnerabilities discovered
each year~\cite{christey2007vulnerability}, and are frequently used 
in real-world attacks~\cite{firehost-superfecta-2013}. 
Additionally, vulnerabilities such as 
Cross Site Request Forgery (CSRF)~\cite{barth2008robust}, command
injection~\cite{su2006essence}, and HTTP response
splitting~\cite{klein2004divide} are also often present in web
applications.

Given such a track record of security problems in both embedded systems and
web applications, it is natural to expect the worse from 
\emph{embedded web interfaces}. 
However, as we discuss further, those vulnerabilities are
neither easy to discover, analyze, and confirm, nor do the vendors perform the
necessary security quality assurance of their firmware images.

\noindent \textbf{Analysis of embedded web interfaces:}
While there are solutions that can be used during the design phase of
the software~\cite{hooimeijer2011fast, samuel2011context,
  livshits2013towards, saxena2011scriptgard}, it is also important to
discover and patch existing vulnerabilities before they are found and
exploited ``in the wild''.
This is possible to do either by
static analysis on their source code~\cite{balzarotti2008saner,
  jovanovic2010static, doupe2011fear, Dahse:rips:ndss14}, or by dynamic analysis 
where their code or web interface is typically
exercised against a number of known attack
patterns~\cite{elie-sp2010-AutoBlackBoxWebTest,
  elie-bh2009-EmbedInterfMassInsec}.

Unfortunately, these techniques and tools can be inefficient or
difficult to use for detecting vulnerabilities inside embedded web
interfaces~\cite{fong2007web,elie-sp2010-AutoBlackBoxWebTest}.  
For instance, performing static analysis on embedded web interfaces seem to be 
a rather simple task once the firmware has been unpacked.
One main limitation of this approach is that the web interfaces often rely 
on various technologies (e.g., PHP, CGIs, custom server-side languages).
However, the static analysis tools are usually designed for a particular 
technology, and many static tools are often concentrated around some 
trendy environment (e.g., PHP) leaving the others ``uncovered''. 
In addition to this, 
though sound static analysis tools exist, many other static analysis 
tools are merely ``glorified greps'' and have a large number of \emph{false positives (FP)}, 
which make them problematic to reliably use in an 
automated large scale study. 
On the other hand, dynamic analysis tools~\cite{felmetsger2010toward, huang2003web}
are more generic as
they are less sensitive to the server-side language. Nevertheless, they
require the system or the web interface to be functional.  Unfortunately, it is
challenging to create an environment that can perfectly emulate firmware images for
a broad range of devices based on a variety of computing architectures and 
hardware designs.

\noindent \textbf{Scalable dynamic analysis of embedded web interfaces:}
The easiest way to perform dynamic analysis is to perform it on a live
device. However, acquiring devices to dynamically analyze them is expensive and does not 
scale. At the same time, it is ethically questionable, if not illegal, to test devices 
one does not own (e.g., devices on the Internet).  
Another option is to
extract the web interface files 
from a device and
load them to a test environment, like an Apache web
server. Unfortunately, a large majority of the embedded web interfaces
use native CGIs, bindings to local architecture-dependent tools or 
custom web server features which cannot be \emph{easily} reproduced in 
a different environment (Section~\ref{sec:technique-hosted}). 

Emulating the firmware is an elegant method to perform dynamic
analysis of an embedded system, since it does not require the physical device to be
present and can be completely performed in a controlled
environment while being easy to scale. But emulation of unknown devices is not easy 
because an embedded firmware expects specific hardware to be fully present, such as
peripherals or memory layouts. 
Previous attempts were made at improving emulation of firmware images
by forwarding hardware I/O or \texttt{ioctl} to the
hardware~\cite{zaddach:ndss14,prospect}. 
    These techniques achieve a rather good emulation, but
require the presence of the original device and a great deal of manual setup, 
which does not scale. We noticed that in Linux-based embedded systems 
the interaction with the hardware is usually performed from the kernel. 
Moreover, the web interfaces often do not interact with the hardware or 
this interaction is indirect.

\subsection{Overview of our Approach}
\label{sec:overview}

\begin{figure*}
  \center
  \includegraphics[width=0.90\textwidth]{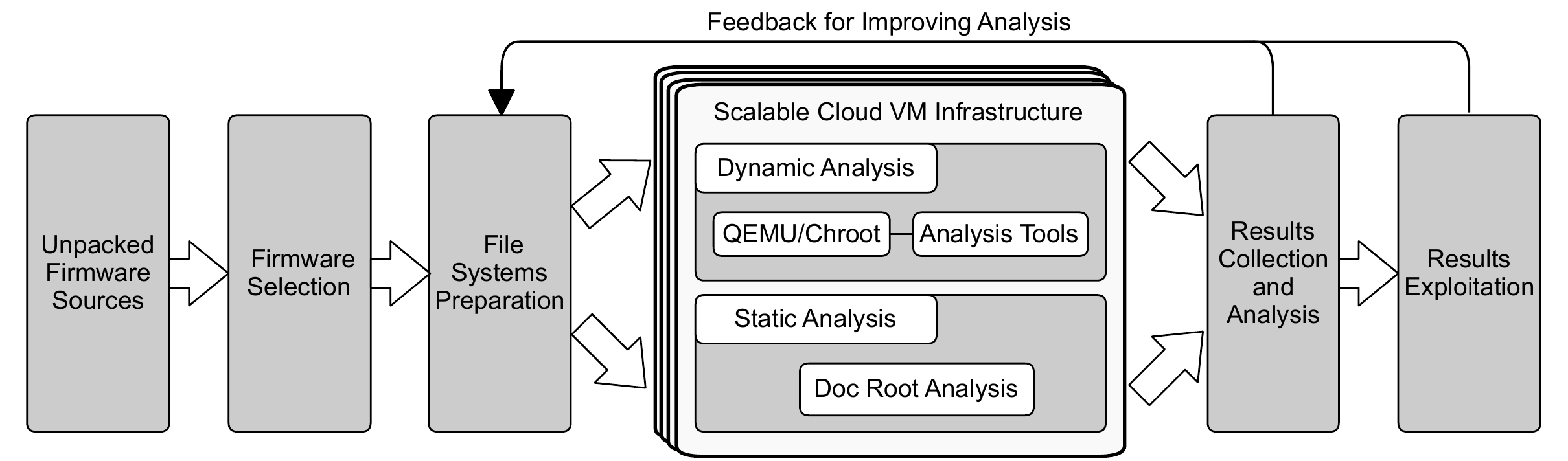}
  \caption{Overview of the analysis framework.}
  \label{fig:overview}
\end{figure*}

To perform scalable security testing of embedded web interfaces we developed a
distributed framework for automated analysis (Figure~\ref{fig:overview}) and 
we tested it in a cloud setup.  We started
our analysis with a dataset of \countfirmwarewebTP{} 
unpacked firmware images
that contain embedded web interfaces.\footnote{We focused mainly on Linux-based 
firmware images. Linux-based firmware images 
are in general well structured and documented, therefore they are easier to 
unpack, analyze and emulate. However, our approach can be easily extended 
in the future to firmware based on system that are similarly well structured 
into bootloaders, kernels and filesystems (e.g., VxWorks, QNX). 
Monolithic firmware is more challenging to fully emulate and in general 
requires additional frameworks, such as Avatar~\cite{zaddach:ndss14}. } 
Then, for each unpacked firmware we identify any potential web
document root present inside the firmware.  At this point we make a
pass with static analysis tools on the modules of the service under test~\footnote{In the particular case of this study, the modules are the files within the \emph{web document root}.}.
Next, we propose a partial emulation of firmware images by replacing
their kernel with a stock kernel (targeting the same architecture) and
emulating the whole userland of the firmware using the QEMU~\cite{qemu}
emulator. We currently support the most frequent architectures that are well supported in QEMU and plan to extend later to other architectures.
We then \texttt{chroot} the unpacked
firmware and start the \texttt{init} program, the init scripts or
sometimes directly the web server. Once (and if) the service under test is
up and operational~\footnote{In the particular case of this study, the service under test is the \emph{web interface}.}, 
we perform dynamic analysis on it. 
Finally, we analyze the results, and whenever applicable we perform manual 
analysis and investigate the failures.

\subsection{Contributions}
\label{sec:contrib}

In this paper we present a completely automated framework to 
perform scalable dynamic firmware analysis. We demonstrate its effectiveness 
by testing the security of embedded web interfaces. Our framework mainly 
relies on the emulation of the firmware images. This allows to test the 
embedded web interfaces using off-the-shelf dynamic analysis tools. 

\noindent In summary, we make the following main contributions:

\begin{itemize}

\item We present the first framework that achieves \emph{scalable and automated 
dynamic analysis of firmwares}, and that was precisely developed to 
discover vulnerabilities in embedded devices using the software-only approach. 

\item We highlight the challenges in emulating the firmware images and testing the web interfaces of 
 embedded systems, and describe the techniques that can be used for such tasks. 

\item We describe our framework which leverages multiple techniques and
  state of the art tools.

\item We perform the first large scale, comprehensive, security study on
  web interfaces of embedded systems.

\item We automatically discover \CountDynVulnsTotalHigh{} previously
  unknown serious vulnerabilities in \CountFirmwareDynVulnTotalHigh{} firmware
  images.


\end{itemize}

\subsection{Outline}
\label{sec:outline}

The remainder of this paper is organized as follows. 
In Section~\ref{sec:opti-stat-dynam}, we explore techniques to emulate and 
analyze embedded firmware and their web interfaces. 
In Section~\ref{sec:framework-details}, we present our framework. 
In Section~\ref{sec:dataset}, we describe our dataset.
In Section~\ref{sec:case-studies}, we showcase our results and analyze
several case-studies. 
In Section~\ref{sec:discussions}, we discuss ethical aspects of our work 
and its limitations, as well as we propose solutions to them. 
We summarize the state of the art in Section~\ref{sec:related-work} and 
then conclude with Section~\ref{sec:conclusion}.

\section{Exploring Techniques to Analyze Embedded Web Interfaces}
\label{sec:opti-stat-dynam}

In this section, we summarize the different possibilities for static
or dynamic analysis of embedded web interfaces, their limitations, and 
motivate our final choices.

\subsection{Static Analysis}
\label{sec:static-analysis}

There are many practical advantages to static analysis tools; they are
often automated and do not require setting up too complex test
environments.  In general, they only need the source code (or
application) to be provided to generate an analysis
report.  It is also relatively easy to plug new static analysis tools
for increased coverage or wider support of file formats and source
code languages.  Finally, as a result of all the above, such tools are
scalable and easy to automate.

However, static analysis techniques have well understood limitations.  On the
one hand, they cannot find all the vulnerabilities, 
i.e., \emph{false negatives (FN)}, while
on the other, they also alert on non-vulnerabilities, 
i.e., \emph{false positives (FP)}, 
which becomes increasingly problematic in large scale automated setups. 
Additionally, we found that embedded devices' firmware often rely on
uncommon technologies for which security static analysis tools often
do not exist (e.g., \texttt{lua, haserl, binary CGIs}).
Albeit there exist a number of static analysis tools for the 
PHP language~\cite{jovanovic2006pixy, Dahse:rips:ndss14},
in our dataset only \printpercent{\CountFirmwarePhp}{\countfirmwarewebTP} of embedded firmware images 
contain PHP code in their server-side. This is not really
a surprise since PHP is not primarily designed for embedded systems. We
nevertheless analyze these cases with
\emph{RIPS}~\cite{Dahse:rips:ndss14} in Section~\ref{sec:case-php}.
Finally, \emph{binary static analysis} can be applied to binary
CGIs to find vulnerabilities such as buffer overflow, (remote) code execution,
command injection~\emph{In this paper we use ``command injection'' and ``command execution'' terms interchangeably.}
(e.g., \texttt{Firmalice}~\cite{firmalice2015ndss} or \texttt{Weasel}~\cite{Schuster_CCS13}). Also, new techniques start 
to appear that are able to cope with the diversity of CPU architectures found in 
embedded systems~\cite{CrossArchBugSearch2015}.

\subsection{Dynamic Analysis}
\label{sec:dynamic-analysis}

Dynamic analysis---an analysis that relies on testing an application by
running it---has many benefits.  First, dynamic analysis of web
interfaces is mostly independent from the server-side technology that is used.
For instance, the very same tool can test web interfaces that are implemented in PHP, native CGIs
or custom web scripting engines.  
Second, it can be used to
confirm vulnerabilities found in the static analysis phase.
Although there exist many dynamic analysis tools for security testing 
of web applications~\cite{elie-sp2010-AutoBlackBoxWebTest}, 
unfortunately, they often require significant effort to setup (e.g., environment 
setup), and sometimes additional customization such as adding new 
vulnerability modules for scanning, testing or validation.

For this particular study, we selected web penetration tools that are free and open source so that we
can easily adapt and integrate them in our framework as well fix their defects when
needed.  Based on this we selected \emph{Arachni}~\cite{foot-arachni},
\emph{Zed Attack Proxy (ZAP)}~\cite{foot-zap}, and \emph{w3af}~\cite{foot-w3af}
to be used in our framework.

However, our approach and framework are designed in a way that allows 
great flexibility. For example, as depicted in the Figure~\ref{fig:vm_qemu} 
other tools such as Metasploit~\footnote{\url{http://www.metasploit.com/}} 
and Nessus~\footnote{\url{http://www.tenable.com/products/nessus-vulnerability-scanner}} 
can supplement or replace the web penetration tools mentioned above. 
In this way, we can achieve additional security and vulnerability testing 
that can help us increase the surface of vulnerability discovery for both 
known and unknown vulnerabilities.

\subsection{Limitations of Analysis Tools}
\label{sec:discuss-limitationsdyntools}

Our framework relies on existing web analysis tools, which have their
own limitations.  For instance, the number of FPs and FNs of this
study are a direct consequence of the vulnerability finding tools we
rely on. An example of their limitations is their ability to detect
\emph{command injections} vulnerabilities.  Those are frequently
missed because such flaws are often hard to discover via automated
testing~\cite{foot-owasp-cmdinj,elie-sp2010-AutoBlackBoxWebTest}.  For
example, tools try to inject commands, such as \texttt{ping <ip>}
assuming that the network is functional and that the targeted system
supports the \texttt{ping} utility.  We overcome some of these
limitations by taking advantage of our ``white box'' approach
(Section~\ref{sec:dyn-our-capt-fs}).

\label{sec:discuss-bugsdyntools}
Additionally, the tools we use were not initially targeting vulnerabilities in
embedded web interfaces and were not designed to be 
integrated in a framework like ours. As a consequence, we found many
problems with those tools which were severely impacting the success
rate of the vulnerability discovery.  We were able to improve or fix
many of them at the cost of a significant engineering effort.
Nevertheless, fixing these bugs proved necessary to obtain better
results.\footnote{We plan to submit the bugfixes to be included in the
  upstream releases of the tools.}  This highlights that better web
application analysis tools are needed, especially ones that are in particular adapted for
testing embedded web interfaces.

\subsection{Running Web Interfaces}
\label{sec:emulating-firmwares}

\begin{figure}
  \center
  \includegraphics[width=0.5\textwidth]{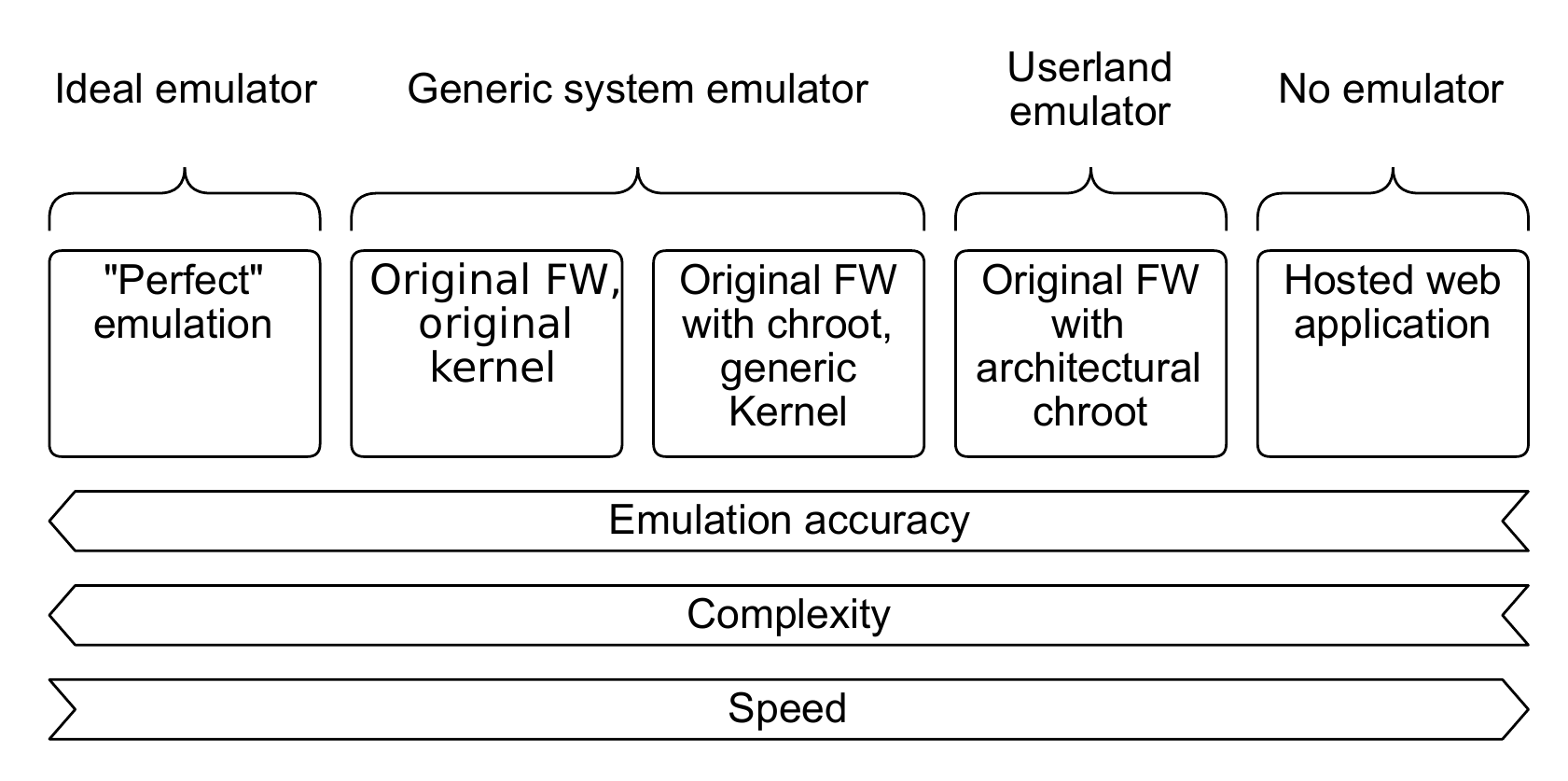}
  \caption{Embedded web interfaces emulation possibilities: from
    perfect emulation of a hardware platform to hosting the web
    interface. (The arrows show a general increasing trend, actual
    evolution of the properties may not be linear.)
  }
  \label{fig:emul_hosting}
\end{figure}

Dynamic analysis of web applications requires a functioning web
interface. There are different ways to launch the web 
interface that is present in the firmware of an embedded system, however,
none of them are perfect. Some methods are very accurate but
infeasible in our setup, such as emulating the firmware in a perfect
emulator (which is not available). Other methods are much less
accurate, like extracting the web application files and serving them
from a generic web server. Therefore, we evaluated different
approaches (Figure~\ref{fig:emul_hosting}) and describe their advantages 
and drawbacks. 

\subsubsection{Hosting Web Interfaces Non-Natively}
\label{sec:technique-hosted}

A straightforward way to launch a web interface from a firmware is to
extract and then launch it under a web server
on an analysis environment, without trying to emulate the original web
server and firmware.
The web application is
located (i.e., the document root, as described in Section~\ref{sec:web-servers-docroots}),
extracted and ``transplanted'' to the \emph{hosting environment}.  
The main advantage of this technique is that it does not require emulation, 
which dramatically simplifies the deployment and thus is easy to automate and scale.

However, this approach has many limitations. For example, it is not possible to handle platform dependent binaries and
CGIs. We analyzed the document roots within 
\countfirmwareforemulation{} firmware candidates for emulation and
found that \PrcntFirmwareBinaryCGI{} out of these were using binary
CGIs or were in some way bound to the platform.\footnote{This is a
  lower bound as we did not count web scripts calling local system
  utilities, e.g., using the \texttt{system()} call.} In addition to
this, the firmware images often use either customized web servers, or
versions which are not available on normal systems, thus a
generic web server (e.g., Apache) has to be used in the hosting environment.
We evaluated this technique and we present results of its evaluation in
Section~\ref{sec:case-compare-host-emu}, where we also compare its performance
to other techniques we use.

\subsubsection{Firmware and Web Interface Emulation}
\label{sec:dynamic-nativeemu}

A preliminary step to emulate a firmware is to know its
architecture. While this may seem straightforward, it is actually
often a complicated step to perform automatically at scale.
For instance, some firmware packages contain files for various
architectures (e.g., ARM and MIPS). Sometimes, vendors package two
different firmware blobs into a single firmware update package. The
firmware installer then picks the right architecture during the
upgrade based on the detected hardware. In such cases, we try to
emulate this filesystem with each detected architecture.  We
detect the architecture of each executable in a firmware using ELF
headers or statistical opcode distribution for raw binaries.  We then
decide on the architecture by counting the number of
architecture specific binaries it contains. Once we detect the right architecture, 
we use the QEMU emulator for that particular architecture. There are
different possibilities for emulating the firmware images, which we now 
compare.

\paragraph{Perfect emulation}

Ideally, the firmware would be complete (including the bootloader,
kernel, etc.)  and a QEMU configuration which perfectly emulates the
original hardware would be available.
However, QEMU only emulates few platforms for each CPU architectures and thus
perfectly emulating unknown hardware is impossible in practice, especially considering that
hardware devices can be arbitrarily complex.
In addition to this, hardware in embedded devices is often custom and
its documentation is, in general, not available. It is therefore infeasible to adapt the
emulator let alone to apply this at a large scale.

\paragraph{Original kernel and filesystem on a generic emulator}

Reusing the kernel of the firmware could lead to a more accurate
emulation, in particular because it may export interfaces for some
custom devices that are needed to properly emulate the system.
Unfortunately, kernels for embedded systems are often customized and
hence do not support a wide range of peripherals. Therefore, using
the original kernel is unlikely to work very well on a generic
emulator.  Additionally, in our dataset only 5\% of the
firmware images were containing the kernel making this approach not
feasible.

\paragraph{Firmware chroot with a generic kernel and filesystem}

Lacking the original kernel, it is is possible to rely on a complete
generic system (for the same CPU architecture of the firmware), which
is then used as a base for the analysis.\footnote{We use the pre-compiled Debian Squeeze packages from~\cite{aurel32qemu}.}
From this generic system
we chroot to the unpacked firmware and execute the shell (e.g.,
\texttt{/bin/sh}) or the init binary (e.g., \texttt{/sbin/init}).
Finally, we start the web server's binary along with the web interface
document root and web configuration.

Ideally, it should be possible to directly boot the firmware filesystem 
instead. However, using a generic file system provides a consistent
environment to control the virtual machine and perform our analysis and monitoring
of the system.
The advantage of this approach is that it allows emulation of the web
interfaces and web server software in their original file system
structure and can execute native programs.

This approach, however, has few drawbacks. First, emulating the system is not very
fast.\footnote{We measured that emulation is one order of magnitude
  slower than native execution.} Additionally, the emulator
environment setup and cleanup introduces a significant overhead. 
Furthermore, with this approach we cannot fully emulate the
peripherals and specific kernel extensions of the embedded
devices. Even so, few firmware images and a limited part of embedded
web interfaces actually interact directly with the peripherals. One
such example is a web page that performs a firmware upgrade which in
turn requires access to \texttt{flash} or \texttt{NVRAM} memory
peripherals.

We found that this approach offers the best trade-off between emulation
accuracy, complexity, and speed (see Figure~\ref{fig:emul_hosting}). It
is also scalable and provided the best results in analyzing dynamically the web 
interfaces (see Section~\ref{sec:case-compare-host-emu}). 

\paragraph{Architectural chroot}
\label{sec:arch-chroot}

One way to improve the performance and emulation management aspects of
our framework is by using \emph{architectural
  chroot}~\cite{foot-arch-root} (also known as \emph{QEMU static
  chroot}). This technique uses \texttt{chroot} to emulate an
environment for architectures other than the architecture of the
running host itself. This basically relies on the Linux kernel's
ability to call an interpreter to execute an ELF executable for a
foreign architecture. Registering the userland QEMU as an interpreter
allows to transparently execute \texttt{ARM} Linux ELF executables on
an \texttt{x86\_64} Linux system. 
However, we found that this approach
was not very stable, making it impossible to use it at a
large scale.  Finally, while this approach has the advantage of
improving emulation speed, in essence, it is unlikely to improve the number of
firmware packages we can finally emulate.  Therefore, we did not
use this technique in our setup, and we leave this for our future work.

\section{Analysis Framework Details}
\label{sec:framework-details}

In order to perform a large scale and automatic analysis of firmware images we
designed a framework to process and analyze them (Figure~\ref{fig:overview}).
First, we obtain a set of unpacked firmware images, analyze and filter them 
(Section~\ref{sec:firmware-selection}). Next, we perform some pre-processing 
of the selected unpacked files. For instance, some firmware images are 
incompletely unpacked or the location of the document root is not obvious 
(Section~\ref{sec:preprocessing-files}). 
We then perform the static and dynamic analyses 
(Section~\ref{sec:analysis-phase}). 
Finally, we collect and analyze the reported vulnerabilities 
(Section~\ref{sec:results-collect-analysis}) and exploit these results 
(Section~\ref{sec:results-exploitation}).

\subsection{Firmware Selection}
\label{sec:firmware-selection}

The firmware selection works as follows. 
First, we select the firmware images that are successfully unpacked and 
are Linux-based systems which we can natively emulate and chroot
(see Section~\ref{sec:preprocessing-files}). 
Second, we choose firmware instances that clearly contain web server binaries 
(e.g., \texttt{httpd, lighttpd}) and typical configuration files 
(e.g., \texttt{boa.conf, lighttpd.conf}). In addition to these, we 
select firmware images that include server side or client side code related to 
web interfaces (e.g., \texttt{HTML, JavaScript, PHP, Perl}).
Our dataset is detailed in Section~\ref{sec:dataset}.

\subsection{Filesystem Preparation}
\label{sec:preprocessing-files}

To emulate a firmware the emulator requires its root filesystem. 
In the simplest case the unpacked firmware 
directly contains the root filesystem. However, in many cases the 
firmware images are packed in different and complex ways. For instance, 
a firmware can contain two root filesystems, one for the upgrade and 
one for the factory restore, or it can be packed in multiple layers 
of archives along with other resources. For these reasons, we first need 
to detect the potential candidates for root filesystems. We achieve this 
by searching for key directories (e.g., \texttt{/bin/, /sbin/, /etc/, /usr/}) 
and for key files (e.g., \texttt{/init, /linuxrc, /bin/sh, /bin/bash, /bin/dash, /bin/busybox}). 
Once we discover such files and folders relative to a directory within 
the unpacked firmware, we select that 
particular directory as the \emph{root filesystem} point. 
There are also cases where it is hard or impossible to detect the root 
filesystem. A possible reason for this is that some firmware updates are just 
partial and do not provide a complete system. 
We extract each detected root filesystem and pack it as a standalone
root filesystem ready for emulation.

Unpacking firmware images can produce ``broken'' root filesystems which we attempt to 
fix. Additionally, in order to start the web server within 
the root filesystem, we need to detect the web server type, its configuration, and 
the document root. 
For these reasons, we have to use heuristics on the candidate root filesystems 
and apply transformations before we can use them for emulation and analysis. 

\subsubsection{Filesystem Sanitization}
\label{sec:filesystem-sanitization}

Unpacking firmware packages is not a perfect procedure. First,
unpacking tools sometimes have defects. Second, some firmware images
have to be unpacked using an imperfect ``brute force''
approach~\cite{costin2014large}.  Finally, some vendors customize
archives or filesystem formats.  For example, some filesystems have 
symbolic links that are incorrectly unpacked because they were represented
as text files containing the target of the link.\footnote{For
example, the symbolic link \texttt{/usr/bin/boa -> /bin/busybox} is
represented with a text file named \texttt{/usr/bin/boa} that
contains the string \texttt{/bin/busybox}.} All these lead to an
incorrect unpacking and thus the unpacked firmware image differs from
the filesystem representation intended to be on the device. This
reduces the chances of successful emulation and therefore we need a
sanitization phase.

This sanitization phase is performed by scripts that traverse unpacked
firmware filesystems and fix such problems. Sometimes, there are
multiple ways to fix a single unpacked firmware. This results in
multiple root filesystems to be submitted for emulation, increasing our
chances of proper emulation of a given firmware. Implementing these 
heuristics added a 
\printpercent{\countrootfsforemulationFixedFS}{\countrootfsforemulationBeforeFixingFS} 
processing overhead. At the same time, it allowed us to increase the successful emulations by 
\printpercent{\CountFirmwareEmulatedOKByFixingFS}{\CountFirmwareEmulatedOK}
and the successful web server launches by 
\printpercent{\CountFirmwareWebServerOKByFixingFS}{\CountFirmwareEmulatedWebServerStarted}.

\subsubsection{Web Server Heuristics}
\label{sec:web-servers-docroots}

Within the firmware, we locate web server binaries and their
related configuration files (e.g., \texttt{boa.conf},
\texttt{lighttpd.conf}).\footnote{Namely:
\texttt{httpd}, \texttt{boa}, \texttt{lighttpd}, \texttt{thttpd},
\texttt{minihttpd}, \texttt{webs}, \texttt{goahead}.} 
The path of the web server and its
configuration file is sufficient to start the web server using a
command such as \texttt{/bin/boa -f /etc/boa/boa.conf} (a real example
from our dataset). Additionally, we extract important settings from
the configuration files (e.g., the document root).

Sometimes, we miss important parameters which are required 
to properly start the web server, such as the document root path or the CGI path.
Often this happens because of a missing configuration file 
(e.g., partial firmware update) or because the parameters are
supplied via the command line from a script which is not available. 
In these cases, we experiment with all the potential document roots of the firmware.
To find a potential document root (within the root filesystem) we first search for index files (e.g.,
\texttt{index.html}, \texttt{default.html}) with possible file
extensions (\texttt{HTML, SHTML, PHP, ASP, CGI}). 
Then, we build a set of \emph{longest common prefix 
directories} of these files. This can result in multiple document
root directories, for example a second document root can be found in 
a recovery partition.
Once we discover the document roots, we prepare the possible commands to start the web server. 
With this, we increase the chances of bringing the web server
up and operational.

We also build an optimized site map for each such document root directory.
We use the site maps to hint the dynamic analysis tools which URLs they have to analyze. 
In general, dynamic analysis tools crawl the web application to discover 
its site map. However, this is inefficient and can easily miss some pages and even whole sets of vulnerabilities~\cite{doupe2010johnny}. 
Thus, we instruct the tools to restrict their analysis 
to the supplied site map and we do this for multiple reasons.
First, it significantly lowers the time required to complete the
dynamic analysis. No time is wasted to analyze uninteresting files,
such as image files, or to (inefficiently) crawl the web application~\cite{doupe2010johnny}.  
Second, it
reduces the chances for the web interface or the emulator to crash by
limiting the resource load, e.g., number of requested files.
Third, it increases the chances that the files that are reported as
vulnerable by static analysis will also undergo dynamic analysis. 

There are several possible improvements to our tests.  Restricting the
site map allows to complete tests in reasonable time but may miss URLs 
when content is dynamically generated or monolithic web 
server binaries are used. Another improvement would be to use a
tool like \texttt{ConfigRE}~\cite{wang2008towards} to automatically
infer configuration files.\footnote{Unfortunately, \texttt{ConfigRE}
  is not available anymore.}

\subsection{Analysis Phase}
\label{sec:analysis-phase}

\begin{figure*}
  \center 
  \includegraphics[width=0.90\textwidth]{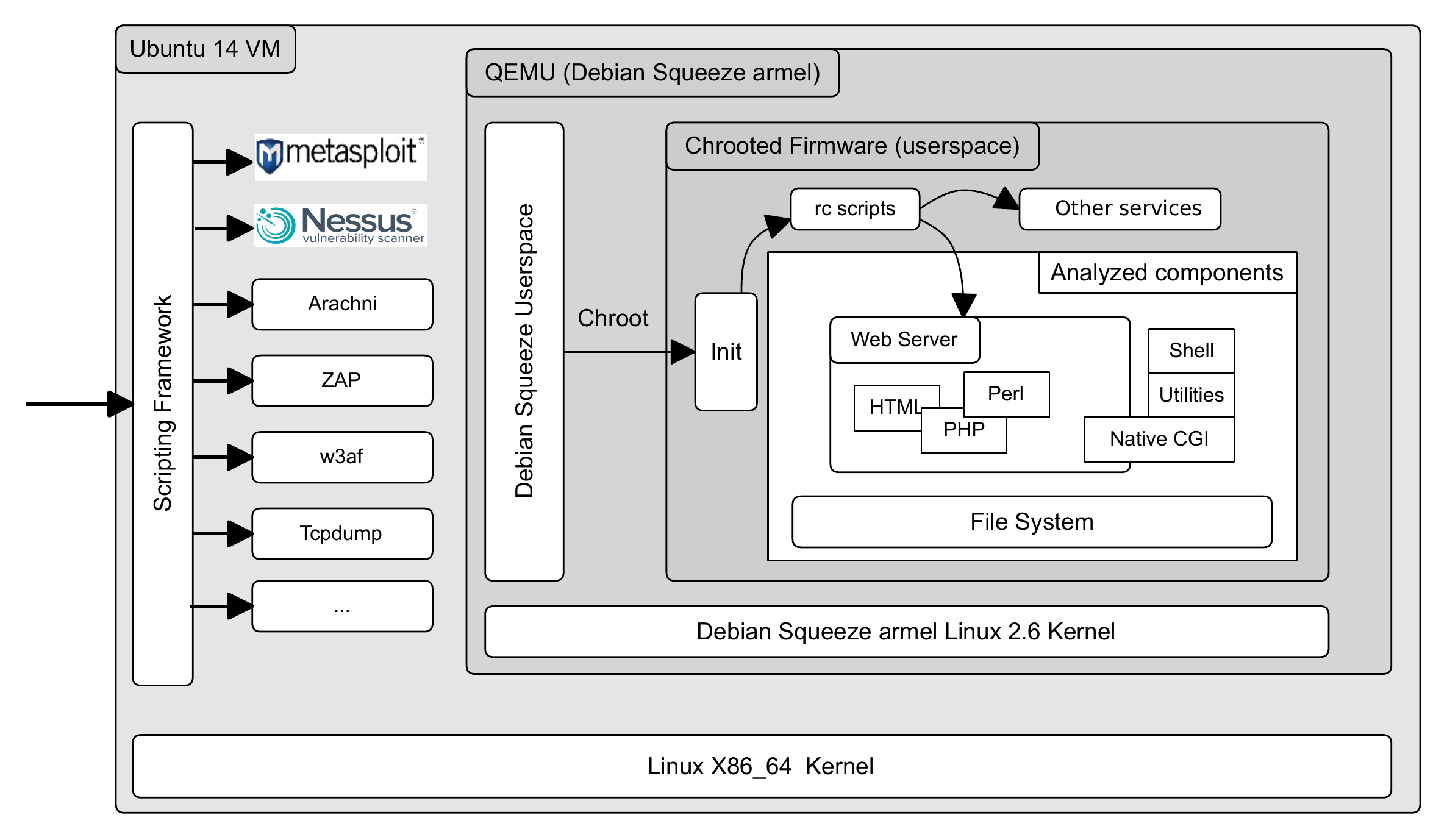}
  \caption{Overview of one analysis environment for Linux armel with a 2.6 kernel.}
  \label{fig:vm_qemu}
\end{figure*}

Once the filesystems are prepared, we emulate each of them in an
analysis Virtual Machine (VM) where dynamic testing is performed
(Figure~\ref{fig:vm_qemu} and Section~\ref{sec:dynamic-analysis}). We
also submit the document roots to the static analyzers
(Section~\ref{sec:static-analysis}). This phase is completely
automated and scales well as each firmware image can be
analyzed independently.

\subsection{Results Collection and Analysis}
\label{sec:results-collect-analysis}

After dynamic and static analysis phases are completed, we obtain the
analysis reports from the security analysis tools. 
We also collect several logs that can help us make
further analysis as well as improve our framework.  These are
typically required to debug the analysis tools or our emulation
environment.  For instance, we collect SSH communication logs with the
emulator host,
changes in the firmware filesystem, and capture the network traffic
of the interaction with the web interface.

\paragraph{File systems changes}
\label{sec:dyn-our-capt-fs}
We capture a snapshot of the emulated filesystem at several different 
points in time. We do this $(i)$~before starting the emulation, 
$(ii)$~after emulation is started, 
and $(iii)$~after dynamic analysis is completed. 
Then, we perform a filesystem \texttt{diff} among these snapshots.
Interesting changes are included in both log files and new files. Log files
are interesting to collect in case a manual investigation is needed.
New files can be the consequence of a \emph{OS command injection} or
more generally of a \emph{Remote Code Execution (RCE)} vulnerability
triggered during the dynamic analysis phase. This often occurs when
dynamic testing tools try to inject commands (e.g., \texttt{touch
  <filename>}). Sometimes, the command injection can be successful but not
detected by the analysis tools. However, it is easy to detect such cases with the filesystem 
diff.

\paragraph{Capturing communications}
\label{sec:dyn-our-capt-tcpdump}

Performing dynamic analysis involves a lot of input and output data 
between the (emulated) device and the dynamic analysis tool. 
Capturing the raw input and output of the communication allows to
increase accountability in case of emulation problem.

For instance, a successful \emph{OS command injection} can go undetected by the 
tools. Also, such vulnerability can be difficult to verify, 
even in a ``white box'' testing approach (Section~\ref{sec:discuss-limitationsdyntools}).
Once the testing phase is over,
it can be discovered that 
a command injection was, in fact, successful. 
In such case, we need to rewind through all HTTP transactions
to find the input triggering the particular vulnerability and afterward we can
look for incriminating inputs and parameters (e.g., a \texttt{touch} 
command).

The testing tools often behave like fuzzers as they try many malformed 
inputs one after the other. Because of this, a detected vulnerability
may not be a direct result of the last input. For example, it can be a
result of the combination of several previous inputs.  It is therefore
important to recover all these previous inputs in order to successfully reproduce
the vulnerability.

\subsection{Results Exploitation}
\label{sec:results-exploitation}

After collecting all the details of the analysis phase, we perform several
steps to exploit these results.  
First, we validate the high impact vulnerabilities by hand and try to create a
proof-of-concept exploit. This could be fully automatized in the future, as
was done for other fields of vulnerability research~\cite{avgerinos2011aeg}. 
Unfortunately, none of the tools we currently use provide such functionality. 
Additionally, from the static analysis reports we manually select the high 
impact vulnerabilities (e.g., command injection, XSS, CSRF) 
and the files they impact.  
We then use these to explicitly drive the dynamic analysis tools and
aim mainly at two things: $(i)$~get the dynamic analysis tools to find
the vulnerabilities they missed (if they did) and $(ii)$~find the
bugs or limitations that prevented the dynamic tools to discover these vulnerabilities in the
first place. 
Even though manual analysis does not scale, it can help uncover additional nontrivial vulnerabilities (see
Table~\ref{tbl:vulns-manual}).
Finally, we summarize all our findings in vulnerability reports to be
submitted as CVEs. 

\section{Dataset}
\label{sec:dataset}

We started with a set of firmware images that we collected over time 
from publicly available sources. 
Table~\ref{tbl:dataset-fw-rootfs} presents details about the counts of 
the firmware images each of the phases in our framework. 

    First, we chose the firmware instances which were successfully unpacked 
and which were Linux-based embedded systems (\countfirmwarewebTP{}). 
These were the systems which seem the easiest to emulate.
Then, we selected firmware instances that clearly contained a web server binary
(e.g., \texttt{httpd, lighttpd}) and typical configuration files 
(e.g., \texttt{lighttpd.conf, boa.conf}). In addition to these, we also 
chose firmware images that included server-side or client-side code associated with 
web interfaces (e.g., \texttt{HTML, JS, PHP, CGI}). 
    Once we applied all the heuristics to the firmware candidates 
(\countfirmwareforemulation{}), we tried to chroot to them and start 
their web interface emulation. 
    Unfortunately, we were able to chroot to only a part of the firmware 
candidates (\CountFirmwareEmulatedOK{}). Then, we were able to start the 
embedded web interfaces only for a part of the firmware images which 
successfully chrooted (\CountFirmwareEmulatedWebServerStarted{}). Finally, 
we were able to discover high impact vulnerabilities only in a part of 
all the web interfaces that were successfully emulated (\countfirmwarestotal{}).

\noindent \textbf{Challenges and Limitations:}
Inevitably, our dataset and the heuristics we apply lead to a bias. 
    First, it almost only contains firmware images that are publicly available 
online. 
    Second, Linux-based devices only account for a portion of all embedded systems. 
    Third, because we use pre-compiled Debian Squeeze images from~\cite{aurel32qemu} 
we performed our tests mainly on ARM, MIPS and MIPSel firmware images. 
However, as we present in Table~\ref{tbl:dataset-archs}, adding support 
for additional architecture should be straightforward and requires mainly 
engineering effort. For example, when pre-compiling the Debian for 
architectures in Table~\ref{tbl:dataset-archs} and using mainline QEMU 
version with additional patches, our framework could ideally support the 
emulation of $\approx$ 97\% of the firmware images in our dataset. 
    Finally, there exist firmware images 
running as monolithic software or embedding web servers which we currently
do not detect or support. 
We are aware of this bias and the results herein should be
interpreted without generalizing them to all embedded systems. 
    In essence, these choices were needed to perform this study and it will be
an interesting future work to extend the study to more diverse
firmware images.

\begin{table}[t]
\centering
\caption{Number of firmware images and corresponding vendors at each phase of the experiment.}
\begin{tabular}{lrrr}
\\
\toprule

\textbf{Dataset phase} & \begin{tabular}{@{}c@{}} \textbf{\# of FWs} \\ \textbf{(unique)} \end{tabular} &   \textbf{\# of root FS}  &   \begin{tabular}{@{}c@{}} \textbf{\# of vendors} \\ \textbf{(unique)} \end{tabular}  \\
\midrule

\textbf{Original dataset}                 &   \countfirmwarewebTP{}           &   --     &  \CountFirmwareWebTPVendors{}   \\
\begin{tabular}{@{}l@{}} Candidates for chroot \\ and web interface emulation  \end{tabular}     &   \countfirmwareforemulation{}    &   \countrootfsforemulationBeforeFixingFS{}    &  \CountFirmwareForEmulationVendors{}    \\
Improved by heuristics                      &   \countfirmwareforemulation{}    &   \countrootfsforemulation{} &    \CountFirmwareForEmulationVendors{}    \\ 
Chroot OK               &   \CountFirmwareEmulatedOK{}          &   --   &   \CountVendorsEmulatedOK{}    \\
Web server OK              &   \CountFirmwareEmulatedWebServerStarted{}    &   --   &   \CountVendorsEmulatedWebServerStarted{}    \\
\begin{tabular}{@{}l@{}} High impact vulnerabilities \\ (static + dynamic) \end{tabular}   &   \countfirmwarestotal{}                      &   --  &   \CountVulnsVendorsTotal{}   \\

\bottomrule

\end{tabular}
\label{tbl:dataset-fw-rootfs}
\end{table}

\begin{table}[t]
\centering
\caption{
    Distribution of CPU architectures, QEMU support of those CPUs, and the 
    success rates of chroot and web launch for each architecture. 
    (The failure analysis is detailed in Section~\ref{sec:results-failures}.) 
}
\begin{tabular}{llrrr}
\\
\toprule

\textbf{Arch.} & \begin{tabular}{@{}c@{}} \textbf{QEMU} \\ \textbf{support} \end{tabular} & \begin{tabular}{@{}c@{}} \textbf{Original} \\ \textbf{firmware} \end{tabular} & \textbf{Chroot OK} & \textbf{Web server OK} \\
\midrule

ARM         &   mainline     &   \printpercent{\CountFWArchArmel}{\countfirmwarewebTP}    &   \printpercent{\CountFWArchArmelEmulationOk}{\CountFirmwareEmulatedOK}    &   \printpercent{\CountFWArchArmelWebserverOk}{\CountFirmwareEmulatedWebServerStarted}    \\
MIPS        &   mainline     &   \printpercent{\CountFWArchMips}{\countfirmwarewebTP}    &   \printpercent{\CountFWArchMipsEmulationOk}{\CountFirmwareEmulatedOK}    &   \printpercent{\CountFWArchMipsWebserverOk}{\CountFirmwareEmulatedWebServerStarted}    \\
MIPSel      &   mainline     &   \printpercent{\CountFWArchMipsel}{\countfirmwarewebTP}    &   \printpercent{\CountFWArchMipselEmulationOk}{\CountFirmwareEmulatedOK}    &   \printpercent{\CountFWArchMipselWebserverOk}{\CountFirmwareEmulatedWebServerStarted}    \\
Axis CRIS   &   patch~\cite{iglesiascris, iglesiasstatus}   &   \printpercent{\CountFWArchCris}{\countfirmwarewebTP}    &   --   &   --  \\
bFLT        &   mainline   &   \printpercent{\CountFWArchBflt}{\countfirmwarewebTP}    &   --    &   --   \\
PowerPC     &   mainline     &   \printpercent{\CountFWArchPowerpc}{\countfirmwarewebTP}    &   --    &   --   \\
Intel 80386 &   mainline     &   \printpercent{\CountFWArchIntel}{\countfirmwarewebTP}    &   --    &   --   \\
DLink Specific &   no   &   $\approx$ \printpercent{\CountFWArchDLinkDirElfMSB}{\countfirmwarewebTP}    &   --    &   --   \\
Unknown     &   no      &   $\approx$ \printpercent{\CountFWArchUnknown}{\countfirmwarewebTP}    &   --    &   --   \\
Altera Nios II &   patch~\cite{wulffaltera}    &   $\ll$ 1\%    &   --    &   --   \\
ARC Tangent-A5 &   no   &   $\ll$ 1\%    &   --    &   --   \\
\midrule

\textbf{Total} & -- &   \textbf{\countfirmwarewebTP} & \textbf{\CountFirmwareEmulatedOK} & \textbf{\CountFirmwareEmulatedWebServerStarted} \\

\bottomrule

\end{tabular}
\label{tbl:dataset-archs}
\end{table}

\begin{table}[t]
\centering
\caption{Distribution of web server types among the \CountFirmwareEmulatedWebServerStarted{} started web server.}
\begin{tabular}{lr}
\\
\toprule

\textbf{Web server} & \textbf{\% among started web servers} \\
\midrule

minihttpd   &   \printpercent{\CountFirmwareEmulatedWebServerStartedMinihttpd}{\CountFirmwareEmulatedWebServerStarted}    \\
lighttpd    &   \printpercent{\CountFirmwareEmulatedWebServerStartedLighttpd}{\CountFirmwareEmulatedWebServerStarted}    \\
boa         &   \printpercent{\CountFirmwareEmulatedWebServerStartedBoa}{\CountFirmwareEmulatedWebServerStarted}    \\
thttpd      &   \printpercent{\CountFirmwareEmulatedWebServerStartedThttpd}{\CountFirmwareEmulatedWebServerStarted}    \\
empty banner      &   26\%    \\

\bottomrule

\end{tabular}
\label{tbl:dataset-web-server}
\end{table}

\begin{table}[t]
\centering
\caption{Web technologies used by the started web servers (combinations possible).}
\begin{tabular}{lr}
\\
\toprule

\textbf{Web interface contains} & \textbf{\% of started web servers} \\
\midrule

HTML        &   \printpercent{\CountFirmwareWebTechHtml}{\CountFirmwareEmulatedWebServerStarted}    \\
CGI         &   \printpercent{\CountFirmwareWebTechCgi}{\CountFirmwareEmulatedWebServerStarted}    \\
PHP         &   \printpercent{\CountFirmwareWebTechPhp}{\CountFirmwareEmulatedWebServerStarted}    \\
Perl        &   \printpercent{\CountFirmwareWebTechPerl}{\CountFirmwareEmulatedWebServerStarted}    \\
POSIX shell &   \printpercent{\CountFirmwareWebTechShell}{\CountFirmwareEmulatedWebServerStarted}    \\

\bottomrule

\end{tabular}
\label{tbl:dataset-web-tech}
\end{table}

\section{Results and Case Studies}
\label{sec:case-studies}

\subsection{Summary of Discovered Vulnerabilities}
\label{sec:results-overview}

Our automated system performed
both static and dynamic analysis of embedded web interfaces inside 
\countfirmwarewebTP{} firmware images from \CountFirmwareWebTPVendors{} vendors.
We found serious vulnerabilities in at least 
\CountFirmwareDynVulnTotalHigh{} firmware images out of 
those \CountFirmwareEmulatedWebServerStarted{} for which we were able to 
emulate the web server. 
These include \CountDynVulnsTotalHigh{} \emph{high impact} vulnerabilities 
found and verified by dynamic analysis. 
Static analysis reported \countfirmwarephpfwripsvuln{} unique firmware 
packages to expose \countfirmwarephpripstotal{} possible vulnerabilities. 
Aggregating static and dynamic analysis reports, a total of 
\countfirmwarestotal{} firmware 
images are responsible for \CountVulnsConclusion{}
vulnerabilities, affecting nearly a quarter of vendors in our 
dataset.\footnote{Some firmware images contribute to both static and dynamic firmware counts.}

\subsection{Static Analysis Vulnerabilities}
\label{sec:case-php}

PHP is one of the most used server-side web programming 
languages~\cite{php-usage}. 
Over the past years, many researchers focused on investigating vulnerabilities
in PHP applications and creating static analysis tools~\cite{Dahse:rips:ndss14, jovanovic2006pixy}.
However, to the best of our knowledge, we are the first to  
study the prevalence of PHP in embedded web interfaces and their security. 
In our dataset the \printpercent{\CountFirmwarePhp}{\countfirmwarewebTP} of embedded firmware images 
contain PHP code in their server-side.
    We extracted the PHP source code from those firmware packages and analyzed the code using RIPS. 
RIPS reported \countfirmwarephpfwripsvuln{} unique 
firmware packages to contain at least one vulnerability and a total of
\countfirmwarephpripstotal{} reported issues. 
The detailed breakdown is presented in 
Table~\ref{tbl:vulns-php}. We observe that cross-site scripting and file manipulation
constitute the majority of the discovered vulnerabilities, while command injection 
(one of the most serious vulnerability class) ranks third.

\begin{table}[t]
\centering
\caption{
  Distribution of PHP vulnerabilities reported by RIPS static analysis. 
  (The typical error rates of each type of vulnerability
  can be found in~\cite{Dahse:rips:ndss14}.) 
}
\begin{tabular}{lrr}
\\
\toprule

\textbf{Vulnerability type} & \textbf{\# of issues} & \textbf{\# of affected FWs} \\ 
\midrule

Cross-site scripting    & \countfirmwarephpripscrosssitescripting    & \countfirmwarefilesphpripscrosssitescripting     \\
File manipulation       & \countfirmwarephpripsfilemanipulation      & \countfirmwarefilesphpripsfilemanipulation       \\
Command execution       & \countfirmwarephpripscommandexecution      & \countfirmwarefilesphpripscommandexecution       \\
File inclusion          & \countfirmwarephpripsfileinclusion         & \countfirmwarefilesphpripsfileinclusion          \\
File disclosure         & \countfirmwarephpripsfiledisclosure        & \countfirmwarefilesphpripsfiledisclosure         \\
SQL injection           & \countfirmwarephpripssqlinjection          & \countfirmwarefilesphpripssqlinjection           \\
Possible flow control   & \countfirmwarephpripspossibleflowcontrol   & \countfirmwarefilesphpripspossibleflowcontrol    \\
Code execution          & \countfirmwarephpripscodeexecution         & \countfirmwarefilesphpripscodeexecution          \\
HTTP response splitting & \countfirmwarephpripshttpresponsesplitting & \countfirmwarefilesphpripshttpresponsesplitting  \\
Unserialize             & \countfirmwarephpripsunserialize           & \countfirmwarefilesphpripsunserialize            \\
POP gadgets             & \countfirmwarephpripspopgadgets            & \countfirmwarefilesphpripspopgadgets             \\
HTTP header injection   & \countfirmwarephpripsheaderinjection       & \countfirmwarefilesphpripsheaderinjection        \\

\midrule

\textbf{Total}          & \textbf{\countfirmwarephpripstotal{}}                 & \textbf{\countfirmwarephpfwripsvuln{} (unique)} \\

\bottomrule

\end{tabular}
\label{tbl:vulns-php}
\end{table}

\subsection{Dynamic Analysis Vulnerabilities}

\begin{table}[t]
\centering
\caption{
  Distribution of dynamic analysis vulnerabilities. 
  (The vulnerabilities
  followed by a ``$\dag$'' sign have low severity and are known to be reported
  with a very high false positive rate, therefore they are not mentioned elsewhere
  in this paper, including when we mention a total number of vulnerabilities found.) 
  }
\begin{tabular}{lrr}
\\
\toprule
\textbf{Vulnerability type} & \textbf{\# of issues} & \textbf{\# of affected FWs} \\
\midrule

\emph{Command execution}        & \emph{\CountDynVulnsCmdInj{}}     & \emph{\CountFirmwareDynVulnCmdInj{}}   \\
\emph{Cross-site scripting}     & \emph{\CountDynVulnsXSS{}}        & \emph{\CountFirmwareDynVulnXSS{}}      \\
\emph{CSRF}                     & \emph{\CountDynVulnsCSRF{}}       & \emph{\CountFirmwareDynVulnCSRF{}}     \\

\midrule
\emph{Sub-total HIGH impact}          & \emph{\CountDynVulnsTotalHigh{}}      & \emph{\CountFirmwareDynVulnTotalHigh{} (unique)}  \\
\midrule

Cookies w/o HttpOnly      $\dag$      & \countdynvulncookiehttp                    & \countfirmwaredynvulncookiehttp                            \\
No X-Content-Type-Options $\dag$      & \countdynvulnxcontenttype                    & \countfirmwaredynvulnxcontenttype                    \\
No X-Frame-Options  $\dag$            & \countdynvulnxframeoptions                    & \countfirmwaredynvulnxframeoptions                      \\
Backup files  $\dag$                  & \CountDynVulnsBackup                    & \CountFirmwareDynVulnBackup                      \\
Application error info $\dag$         & \CountDynVulnAppErrorDiscl                    & \CountFirmwareDynVulnErrorDiscl                      \\

\midrule
Sub-total low impact $\dag$           & \CountDynVulnsTotalLow{}      & \CountFirmwareDynVulnTotalLow{} (unique)  \\
\midrule

\textbf{Total}          & \textbf{\CountDynVulnsTotal{}}      & \textbf{\CountFirmwareDynVulnTotal{} (unique)}  \\

\bottomrule
\end{tabular}
\label{tbl:vulns-dynamic}
\end{table}

Our framework was able to perform dynamic security testing on
\CountFirmwareEmulatedWebServerStarted{} distinct web interfaces, and
the general results are presented in Table~\ref{tbl:vulns-dynamic}.
In particular, we discovered \CountFirmwareDynVulnCmdInj{} firmware
packages which are vulnerable to command injection.  The impact of
such vulnerabilities can be significant as a large number of devices
may be running these firmware images (e.g.,
Section~\ref{sec:discuss-casestudy-cmdinj}).

Additionally, we found that \CountFirmwareDynVulnXSS{} firmware packages
are affected by XSS and \CountFirmwareDynVulnCSRF{} are vulnerable to CSRF. 
Even though XSS and CSRF 
are usually
not considered to be critical vulnerabilities, they can have a high
impact.  For example, Bencs{\'a}th et al.\cite{bencsath2011xcs} 
were able to
completely compromise an embedded device only by using XSS and CSRF
vulnerabilities.  

The above 
vulnerabilities affect the firmware of multiple type of 
devices in our dataset, such as SOHO routers, CCTV cameras, 
smaller WiFi devices (e.g., SD-cards). 
Leveraging tools such as Shodan~\cite{shodan} or ZMap~\cite{zmap13}, 
it is possible to correlate these firmware images to populations of online
devices using multiple correlation techniques~\cite{costin2014large}, 
which we leave for future work. 

In summary, we found vulnerabilities in roughly one out of four 
(\printpercent{\CountFirmwareDynVulnTotal}{\CountFirmwareEmulatedWebServerStarted}) 
of the dynamically tested firmware images, 
which demonstrates the viability of our approach.

\subsection{Evaluation of Hosting Web Interfaces}
\label{sec:case-compare-host-emu}

\emph{``Hosting''} embedded web interfaces seems a promising approach
as it permits testing a web interface without emulating the complete
firmware. Indeed, many firmware images are difficult (or impossible)
to emulate. We therefore tested the ``hosting'' approach on all the firmware images
from our dataset where our web server heuristics tools could extract a document root (Section~\ref{sec:web-servers-docroots}). The document
roots are then ``transplanted'' into testing hosts containing a generic web 
server,\footnote{Ubuntu 14.10, Kernel 3.13, Apache2, PHP 5.5.9, Perl 5.18.2.} 
and then the dynamic analysis is performed with the same tools
as for the emulated firmware images.

Table~\ref{tbl:compare-host-emu} shows the high impact vulnerabilities found
in this experiment and also presents a comparison with the emulation approach.  We can
immediately see that unsurprisingly the ``hosted'' analysis allows to test web interfaces from many more 
firmware images, but surprisingly it almost only reports CSRF
vulnerabilities. In fact the technique did not allow to detect any new command injection or XSS vulnerabilities.
We expect that the lack of results for some categories of vulnerabilities
is due to the fact that using the ``hosted'' approach
with a static web server configuration has some undesired
side effects. For instance, the headers of the HTTP responses will be
different from those of the real device's web server, while these headers may 
have an important security role (e.g., Cookies w/o HttpOnly, No
X-Content-Type-Options, No X-Frame-Options). In fact, for command execution 
and XSS vulnerabilities we had to perform manual interventions into 
the hosting environment to make the web interface more functional. Then we had 
to rerun the dynamic analysis to discover a part of vulnerabilities 
already reported by the fully automated emulation approach. 
In few instances we had to install additional \texttt{apache2} modules, while 
in some others we had to disable \texttt{.htaccess} configuration 
files which came with the transplanted document roots. 
Yet in several other cases we had to adjust a variety of \texttt{shebang (\#$!$)} 
paths in the interpreted scripts' headers to point to the correct interpreter 
path of the hosting environment. 
These manual interventions clearly do not scale and limit the ``hosted'' 
approach. In the future, we plan to address such limitations with 
approaches to automatically reconfigure the ``hosting'' environment 
based on the semantics of the transplanted document root. 
Overall, we conclude that both firmware emulation and ``hosting'' web
interfaces are useful and complementary techniques. Moreover, whenever
the emulation is possible, it finds more vulnerabilities.

\begin{table}[tbh]
\centering
\caption{
  Comparison of firmware images affected by high impact vulnerabilities found 
  with the \emph{Emulated and Chroot} method and the ones found with 
  the \emph{Hosted} technique. 
  The firmware and vulnerabilities marked with a "$\dag$" are found using the \emph{Hosted} technique, which is not yet integrated in the fully automated framework.
  Therefore, they are not aggregated elsewhere in this paper, including when we mention a total number of vulnerabilities and affected firmware. 
}
\resizebox{\columnwidth}{!}{
\begin{tabular}{lrrr}
\\
\toprule

\textbf{} & 
\begin{tabular}{@{}c@{}} \textbf{Emulated} \\ \textbf{(unique FWs)} \end{tabular} & 
\begin{tabular}{@{}c@{}} \textbf{Hosted} \\ \textbf{(unique FWs)} \end{tabular} & 
\begin{tabular}{@{}c@{}} \textbf{Hosted Contribution} \\ \textbf{(added unique FWs)} \end{tabular} \\

\midrule
\emph{Command execution}        & \emph{\CountFirmwareDynVulnCmdInj{}}   &  \emph{\CountFirmwareDynVulnCmdInjHosted{} $\dag$}    &   0 \\
\emph{Cross-site scripting}     & \emph{\CountFirmwareDynVulnXSS{}}      &  \emph{\CountFirmwareDynVulnXSSHosted{} $\dag$}       &   0 \\
\emph{CSRF}                     & \emph{\CountFirmwareDynVulnCSRF{}}     &  \emph{\CountFirmwareDynVulnCSRFHosted{} $\dag$}      &   269 \\ 
\midrule

\begin{tabular}{@{}l@{}} \textbf{Total tested FWs} \end{tabular} &  \textbf{\CountFirmwareEmulatedWebServerStarted{}} &   \textbf{\CountFirmwareHostedDynArachni{} $\dag$} &  269 \\
\begin{tabular}{@{}l@{}} \textbf{Total vulnerable FWs} \end{tabular} &  \textbf{\CountFirmwareDynVulnTotalHigh{}} &   \textbf{\CountFirmwareDynVulnTotalHighHosted{} $\dag$} & 262 \\

\bottomrule
\end{tabular}
}
\label{tbl:compare-host-emu}
\end{table}

\subsection{HTTPS and Other Network Services}
\label{sec:case-https}

We also explored how often embedded devices provide HTTPS support.  In
our dataset, nearly
\printpercent{\countfirmwareswithHTTPScerts}{\countfirmwarewebTP} of
the original firmware images contained at least one HTTPS certificate.
This provides a lower bound estimate of firmware images that could
provide a web server with HTTPS support.\footnote{As some devices may
  generate certificates dynamically.} Similarly, around
\printpercent{\CountFirmwareEmulatedWebserverStartedHTTPS}{\CountFirmwareEmulatedWebServerStarted}
of the firmware instances starting an HTTP web server, also started an
HTTPS one.  We also expect this number to be a lower bound
estimate as an HTTPS web server might not start for multiple reasons.
It is unfortunate that so few devices provide HTTPS support.

While in this paper we focus on the security of web interfaces, we
found it interesting to also report on the other network services that
are automatically started during the dynamic analysis. Indeed, these
additional network services which we detected using 
\texttt{netstat}\footnote{Scanning the virtual machine with the NMAP
scanner~\cite{foot-nmap} was both too slow and provided too shallow
results.}  may be vulnerable on their own (e.g.,
TFTP~\cite{foot-cve-tftf}, TR-069~\cite{foot-cve-tr069},
RTSP~\cite{foot-cve-rtsp}, Debug~\cite{foot-cve-debug}).  For this we
compare the \texttt{netstat} output before and after starting the
chroot and the init scripts.  This provides a very precise information
on which program is listening on which port (Table~\ref{tbl:net-svc}).

\begin{table}[tbh]
\centering
\caption{Distribution of network services  opened by
  \CountFirmwaresOtherServices{} firmware instances out of
  \CountFirmwareEmulatedOK{} successfully emulated ones. The last
  entry summarizes the 16 unusual port numbers opened by
  services such as web, telnetd, ftp or upnp servers.
}
\begin{tabular}{lrlr}
\\
\toprule
\textbf{Port type} & \textbf{Port number} & \textbf{Service name} & \textbf{\# of FWs} \\
\midrule

TCP     &   554         &   RTSP            &     91      \\
TCP     &   555         &   RTSP            &     84      \\
TCP     &   23          &   Telnet          &     60      \\
TCP     &   53          &   DNS             &     23      \\
TCP     &   22          &   SSH             &     15      \\
TCP     &   Others      &   Others          &     58      \\

\midrule
\textbf{Total} &   &                  &   \textbf{\CountFirmwaresOtherServices{} (unique)} \\

\bottomrule

\end{tabular}
\label{tbl:net-svc}
\end{table}

\subsection{Analysis of the Failures}
\label{sec:results-failures}

The failures at various stages limit the coverage of the tested 
firmware images. 
For example, Table~\ref{tbl:dataset-archs} shows that chroot failed for around 
\printpercent{\CountFirmwareEmulatedNOTOK}{\countfirmwareforemulation} 
of the original firmware images, and around \printpercent{\CountFirmwareEmulatedWebServerNOTStarted}{\CountFirmwareEmulatedOK} 
of the successfully chrooted firmware packages failed to start the embedded web interface. 
To increase the coverage and hence the chances of finding more 
vulnerabilities in more firmware images, we have to perform analysis of 
the failures and improve our framework. 

Such a \emph{failure diagnosis} is very time consuming~\cite{li2010cost}, 
it requires the exploration of the failure symptoms 
(e.g., message patterns, error codes, unstructured or inconsistent logs). 
However, this information differs from one system (i.e., device, firmware) 
to another. 

Ideally, such fixes should resolve the failures permanently.
However, in practice failures often reoccur ~\cite{lee2000diagnosing}.
One reason is that the corrective maintenance activities, failure diagnosis 
and solution development can take a long effort. Another reason is that 
the deployed solution is not completely effective. 
Additionally, failures can become more recurrent in older systems 
(e.g., old devices and firmware). 
Moreover, once the firmware complexity grows, human analysts become
overloaded with the generated logging or failure information.
Therefore, scalable failure analysis approaches are required

\paragraph{Analysis}

As mentioned before, during the experiments our framework encountered 
\CountFirmwareEmulatedNOTOK{} firmware images where \emph{chroot} failed, and 
\CountFirmwareEmulatedWebServerNOTStarted{} firmware emulations where 
\emph{web interface} launch failed. 
However, this is too many failures to analyze manually, and the 
diversity of the systems makes automated log analysis challenging.
Therefore, we performed the analysis on a sample of the data and 
we apply statistical methods and confidence intervals to reason 
about failures, their root causes and to find ways to improve the system.
    For this, we consider a 95\% confidence level and a $\pm$ 10\% confidence 
interval for the accuracy of estimations. Those parameters allow to provide coarse grained 
results while remaining within a reasonable number of failures to manually analyze.

We analyzed the log samples of \EmulatedNOTOKRandSampl{} randomly
selected (out of \CountFirmwareEmulatedNOTOK{}) firmware files that
\emph{failed to chroot}. Among those we found actual chroot failures
and cases where chroot was successful but which we failed to exploit.

    There were \EmulatedNOTOKRandSamplChrootFail{} cases where chroot 
was actually the cause of the failure (which we extrapolate to \EmulatedNOTOKRandSamplChrootFailPct{} 
$\pm$ \EmulatedNOTOKRandSamplChrootFailConfInt{} cases out of \CountFirmwareEmulatedNOTOK{}). 
In these sampled logs we found two main reasons of chroot failure:

\begin{itemize}

\item Chroot failed for \EmulatedNOTOKRandSamplChrootFailExecFormat{} 
firmware images because of \emph{exec format errors} (extrapolated 
to \EmulatedNOTOKRandSamplChrootFailExecFormatPct{} $\pm$ 
\EmulatedNOTOKRandSamplChrootFailExecFormatConfInt{} out 
of \CountFirmwareEmulatedNOTOK{} firmware images). 
Those failures are the consequence of an incorrect guess of the CPU architecture or 
due to a \texttt{/bin/sh} that contain \emph{illegal instructions}.\footnote{
For example, MIPS processors can have customized opcodes. This is possible 
by using User Defined Instruction Sets (UDIs)~\cite{ienne2006customizable}.}
We believe that those error cases should be relatively \emph{easy} to fix, 
i.e., by changing the QEMU architecture (e.g., because architecture was improperly 
detected in the first place) or improving QEMU (e.g., to support, or ignore, non standard instructions). 

\item  Second, chroot failed for \EmulatedNOTOKRandSamplChrootFailFS{} 
samples because the firmware images were only \emph{partial firmware updates} 
(which we extrapolate to \EmulatedNOTOKRandSamplChrootFailFSPct{} $\pm$ 
\EmulatedNOTOKRandSamplChrootFailFSConfInt{} out of \CountFirmwareEmulatedNOTOK{}).
Such firmware images do not contain any shell or busybox binary that our framework 
could chroot to. 
Those cases can also be fixed by replacing missing utilities, however,
at this point the firmware under analysis will start to diverge from
the actual device's firmware.

\end{itemize}

    The remaining \EmulatedNOTOKRandSamplChrootDetectFail{}  chroot failures were 
found to be \emph{false positives}. In fact, those firmware images 
did  chroot successfully but our framework failed to detect this.
This can occur in systems that set the environment in unusual ways or 
take long time to respond to chroot and environment queries (timeout). 
We therefore extrapolate that \EmulatedNOTOKRandSamplChrootDetectFailPct{} $\pm$ 
\EmulatedNOTOKRandSamplChrootDetectFailConfInt{} of \CountFirmwareEmulatedNOTOK{} 
chroot failure cases were in fact successfully chrooted. 
We estimate that those cases should be relatively easy to fix. The fixes could include more 
adaptive timeouts, and more robust handling of various shells and 
environments. 

    In summary, for the \emph{chroot failed} failures we estimate that 
\EmulatedNOTOKRandSamplFixEASY{} samples should be relatively \emph{easy} to fix, 
meaning that \EmulatedNOTOKRandSamplFixEASYPct{} 
($\pm$ \EmulatedNOTOKRandSamplFixEASYConfInt{}) of the failures should be \emph{easy} to fix  fix
that would allow the emulation to advance one step further. 

Similarly, we analyzed log samples of \WebServerNOTOKRandSampl{} randomly selected 
(out of \CountFirmwareEmulatedWebServerNOTStarted{}) firmware files that 
successfully chrooted but failed to start the web interface. 
    We found \WebServerNOTOKRandSamplDeviceFail{} instances where \emph{missing device} 
were the cause of the failure. Some examples of missing devices are 
\texttt{eth1, br0, /dev/gpio, /dev/mtdblock0}. We estimate that fixing 
the \emph{missing devices} in the emulator is generally \emph{hard}, and 
sometimes even impossible due to the lack of specification sheets. 
This also means that \WebServerNOTOKRandSamplDeviceFailPct{} $\pm$ 
\WebServerNOTOKRandSamplDeviceFailConfInt{} of those 
\CountFirmwareEmulatedWebServerNOTStarted{} web server failures are 
in general \emph{hard} to fix. 
    We also found \WebServerNOTOKRandSamplInitFail{} firmware samples 
that failed or hung during the initialization of the emulation. 
Some of these errors were \emph{``Init is the parent of all processes''} 
and \emph{``init: must be run as PID 1''}. The reason for such errors 
could be the chrooted nature of the emulation. However, we expect 
this not to be too difficult to fix. This translates to \WebServerNOTOKRandSamplInitFailPct{} 
$\pm$ \WebServerNOTOKRandSamplInitFailConfInt{} of original 
\CountFirmwareEmulatedWebServerNOTStarted{} web interface failure 
will be likely \emph{easy} to fix. 
    Finally, we identified \WebServerNOTOKRandSamplWebservFail{} firmware 
samples that reached the web interface launch but failed to launch. 
We therefore estimate that this is the case for \WebServerNOTOKRandSamplWebservFailPct{} $\pm$ 
\WebServerNOTOKRandSamplWebservFailConfInt{} of firmware images that 
produced \emph{web server} failures. Examples of errors are 
\emph{``(server.c.621) loading plugins finally failed''} 
and \emph{``(log.c.118) opening errorlog /tmp/log/lighttpd/error.log failed: No such file or directory''}. 
However, we estimate this failure category can be relatively \emph{easy} to fix. 

    In summary, for the \emph{web server} failures we estimate that 
\WebServerNOTOKRandSamplFixEASY{} samples should be relatively \emph{easy} to fix, 
meaning that \WebServerNOTOKRandSamplFixEASYPct{} 
($\pm$ \WebServerNOTOKRandSamplFixEASYConfInt{}) of the failures should have \emph{easy} 
fixes that would eventually allow the launch of embedded web interfaces.

\paragraph{Further improvements}

The failure analysis and determination in large-scale deployments~\cite{chen2002pinpoint} 
can be improved and automated in several ways. 
One way is to perform log pre-processing~\cite{zheng2009system}, 
log mining~\cite{lim2008log} and analysis~\cite{xu2009detecting}. 
This approach often uses clustering and machine learning techniques 
to classify an unknown execution of the system based on its logs and 
based on previously seen logs of that system~\cite{cohen2004correlating,lin1990error}. 
Filesystem instrumentation is another approach to automate the failure 
analysis~\cite{huang2011assisting}. Such an approach is generic because it 
does not assume that the system is based on particular components or existing 
log files. The failure causes are determined by looking at differences 
between file accesses (e.g., which file, when) under both normal and abnormal 
executions. 
However, these approaches assume that there exist 
samples of the analyzed system that runs under normal and abnormal conditions. 
Also, some of these approaches require domain-specific knowledge~\cite{lim2008log}. 
These techniques are not trivial to apply in our case. First, we aim at 
emulating unknown firmware images regardless of the type or application 
domain of the device for which the firmware is intended. 
Second, most of the times we do not have samples of a non-failure run of 
the firmware. 

Another way to trace and analyze the failures is to use tracing~\cite{ding2011automatic}. 
For example, \texttt{strace} is a debugging tool that provides useful 
diagnostics by tracing system calls and signals. 
Unfortunately, \texttt{strace} is broken for stock kernels 
2.6\footnote{\url{http://landley.net/aboriginal/news.html}}, which also 
affects the builds for embedded architectures (e.g., ARM, MIPS). 

Finally, kernel level instrumentation and analysis could be a reliable 
approach to monitor~\cite{jarboui2002experimental} and detect the failures 
and their root cause. 
For example, \texttt{Kprobes}~\cite{mavinakayanahalli2006probing,krishnakumar2005kernel} 
can be used to dynamically monitor any kernel routine and
collect debugging and performance information 
non-disruptively.\footnote{\url{https://www.kernel.org/doc/Documentation/kprobes.txt}}
Unfortunately, Kprobes is often not enabled in default kernels we used 
but more importantly it's support for various CPU architectures is not stable
(at least in some old kernel versions that we need to use for emulation).

These limitations do not represent themselves research challenges. 
However, it takes more than a trivial engineering effort to address them 
and overcome their effect in a systematic and generic manner. 
We leave the resolution of such limitations as an engineering challenge 
for future work.

\subsection{Case Study: Netgear Networking Devices}
\label{sec:discuss-casestudy-cmdinj}

\newcommand{\NetgearDeviceTypes}{8}

In the results set we discovered many interesting cases. One of them is the 
case of at least \NetgearDeviceTypes{} different device types from \emph{Netgear}, 
one of the major networking equipment manufacturers. 
First, our dynamic analysis framework \emph{automatically discovered} several 
previously unknown yet important \emph{command injection} and \emph{XSS} 
vulnerabilities that can be exploited by non-authenticated 
users~\footnote{The latest firmware versions for some device types now require 
authentication to access the vulnerable modules. However, the modules still 
remain vulnerable to command injection and XSS triggered by authenticated 
users because the root cause problems, i.e., unsafe and unsanitized 
\texttt{exec} and \texttt{echo} calls were not actually fixed.}. 
Then, we also manually verified and confirmed that the discovered 
vulnerabilities are indeed exploitable on the emulated interfaces and 
on some physical devices we acquired for confirmation. 

The affected modules are written in PHP and are used to write (i.e., store 
onto the hardware board) and display back to the user some manufacturer 
data, such as MAC address and hardware registers values. 
These values are supplied by the user and therefore can be controlled by the 
attacker. To write the manufacturer data onto the hardware board, the 
affected modules use the unsafe \texttt{exec} call with unsanitized input 
which leads to \emph{command injection}. To display the manufacturer data 
back to the user the affected modules use the unsafe \texttt{echo} with 
unsanitized input which leads to \emph{XSS}. 
    The PHP modules affected by the discovered vulnerabilities do not have 
a clear purpose because the information they provide can be accessed from other 
pages of the web interface. Also, unless the user or the attacker knows the 
affected PHP modules names, the user cannot reach these modules 
by browsing the web interfaces because no other pages link to them. 
Therefore, one could speculate that these modules, at best, might be 
forgotten debugging files used during development or, at worse, could be 
classified as potential backdoors. 

Interestingly enough, since the affected modules were written in PHP, 
we were also able to discover these vulnerabilities by using the static 
analysis tools such as RIPS. This confirms that our approach 
to use a combination of static and dynamic analysis (Section~\ref{sec:overview}) is 
sound and efficient, and combining them both can drive the dynamic analysis to focus more 
on particular modules that are flagged in the static analysis phase. 
Once again, the disadvantage of the static analysis tools is the high number 
of false positives and the verboseness of the output. For this reason, 
if we perform only the static analysis on these modules we could have missed 
these vulnerabilities easily. 

An additional manual analysis (Table~\ref{tbl:vulns-manual}) revealed
that those devices suffer from some more pre-authentication
vulnerabilities, such as privilege escalation to web admin,
unencrypted configuration storage and unauthorized configuration
downloads (e.g., WPAx keys, passwords).

\begin{table}[t]
\centering
\caption{Distribution of vulnerabilities motivating the manual
  analysis (Section~\ref{sec:results-exploitation}). Firmware
  images relate to similar products of one vendor. 
  (These vulnerabilities were manually found so we don't 
  consider them when we mention a total number of vulnerabilities 
  found by our automated framework.) 
}
\begin{tabular}{lr}
\\
\toprule
\textbf{Vulnerability type} & \textbf{\# of affected FWs} \\
\midrule

Privilege escalation (full admin)     &   \countfirmwaremanualfwprivesc   \\
Unauthorized configuration download   &   \countfirmwaremanualfwunauthconfdown    \\
Unencrypted configuration storage    &   \countfirmwaremanualfwunencryptconf \\

\midrule
\textbf{Total high impact}          &   \textbf{\countfirmwaremanualfwtotal{} (unique)}    \\
\bottomrule

\end{tabular}
\label{tbl:vulns-manual}
\end{table}

Additionally, by using the emulated interfaces we extracted the web interface 
keywords that we used to perform searches on Shodan and Google for 
potentially affected online devices. 
Shodan reported around 500 affected devices, which seem to be a small
population of affected devices connected directly to the Internet via
their WAN interface.
However, many wireless routers are used mainly in WLANs and cannot be found from the Internet. 
The \emph{WIGLE} project provides access to worldwide scans of wireless
networks and can be used to detect devices of this particular vendor.
At the date of this writing, the \emph{WIGLE} project reports that several 
millions~\footnote{\url{https://wigle.net/stats\#ssidstats}} of wireless 
devices from \emph{Netgear} are deployed worldwide. However, lacking detailed 
information of device types in the \emph{WIGLE} database, we cannot exactly 
tell how many actual devices worldwide are affected by these particular 
vulnerabilities.

Finally, the Table~\ref{tbl:netgear-boardData} lists the SHA-256 and the byte size of 
each PHP module that our framework automatically found vulnerable to the 
\emph{command injection} and \emph{XSS} we detailed above. These PHP modules 
can be found in more than 30 firmware firmware images that are available 
in total for the affected \NetgearDeviceTypes{} device types~\footnote{
Following a responsible disclosure practice, we intentionally omitted the 
names of the affected devices and vulnerable modules.}.

\begin{table}[t]
\centering
\caption{
List of the SHA-256 and the byte size of each PHP module that our framework 
automatically found vulnerable to the \emph{command injection} and \emph{XSS} 
in at least \NetgearDeviceTypes{} device types from Netgear. 
}
\begin{tabular}{lr}
\\
\toprule
\textbf{SHA-256} & \textbf{Byte size} \\
\midrule

03bd170b6b284f43168dcf9de905ed33ae2edd721554cebec81894a8d5bcdea5 & 4847 \\ 
2311b6a83298833d2cf6f6d02f38b04c8f562f3a1b5eb0092476efd025fd4004 & 3646 \\ 
325c7fe9555a62c6ed49358c27881b1f32c26a93f8b9b91214e8d70d595d89bb & 4838 \\ 
33a29622653ef3abc1f178d3f3670f55151137941275f187a7c03ec2acdb5caa & 4922 \\ 
35c60f56ffc79f00bf1322830ecf65c9a8ca8e0f1d68692ee1b5b9df1bdef7c1 & 4914 \\ 
40fbb495a60c5ae68d83d3ae69197ac03ac50a8201d2bccd23f296361b0040b9 & 3582 \\ 
453658ac170bda80a6539dcb6d42451f30644c7b089308352a0b3422d21bdc01 & 5039 \\ 
4679aca17917ab9b074d38217bb5302e33a725ad179f2e4aaf2e7233ec6bc842 & 3638 \\ 
56714f750ddb8e2cf8c9c3a8f310ac226b5b0c6b2ab3f93175826a42ea0f4545 & 4166 \\ 
70fe0274d6616126e758473b043da37c2635a871e295395e073fb782f955840e & 3544 \\ 
760bde74861b6e48dcbf3e5513aaa721583fbd2e69c93bccb246800e8b9bc1e6 & 3684 \\ 
8bf836c5826a1017b339e23411162ef6f6acc34c3df02a8ee9e6df40abe681ff & 4964 \\ 
9f56e5656c137a5ce407eee25bf2405f56b56e69fa89c61cdfd65f07bc6600ef & 4256 \\ 
a5ef01368da8588fc4bc72d3faaa20b21c43c0eaa6ef71866b7aa160e531a5b4 & 3791 \\ 
dcefcff36f2825333784c86212e0f1b73b25db9db78476d9c75035f51f135ef6 & 3552 \\ 

\bottomrule

\end{tabular}
\label{tbl:netgear-boardData}
\end{table}

\subsection{Case Study: Samsung CCTV Cameras}
\label{sec:discuss-casestudy-samsung}

Another interesting case study is the one of a Samsung CCTV camera model. 
The affected camera model has Ethernet networking, provides a web interface 
and multiple advanced functions (e.g., face detection and tracking). 
These cameras are intended for SOHO and enterprise setups, and cost several 
hundreds US dollars. 
    Our system \emph{automatically discovered} a \emph{command injection} and 
multiple \emph{XSS} vulnerabilities in their web interfaces as follows. 
    First, our system applied the static analysis. Since a part of the camera's 
web interface consists of CGI scripts written in PHP, the RIPS tool reported 
multipled potential vulnerabilities of different types, including 
\emph{command injection} and \emph{XSS}. 
interface (i.e., the web interface is not exclusively implemented by native 
    Then, our system took advantage of the partial PHP implementation of the web 
binaries) and applied the \emph{Hosted} technique (Section \ref{sec:emulating-firmwares}) 
to transplant the camera's web interface onto a Ubuntu Linux host. 
    Finally, our system used the report from the RIPS tool to focus the dynamic 
analysis phase onto the modules contained in the report. As a result, the dynamic 
analysis tools were able to confirm that one particular 
module~\footnote{The details of the vulnerability can be found at 
\url{http://firmware.re/vulns/}} of the web interface allowed an attacker 
to successfully perform \emph{command injection} and \emph{XSS}. 
    This once again confirms that our approach to use a combination of static and 
dynamic analysis (Section~\ref{sec:overview}) is sound and efficient. Also, 
it also confirms that ``hosting'' the embedded web interfaces non-natively 
is effective. We show it can successfully achieve discovery and confirmation 
of high-severity vulnerability similar to the emulation-based approaches.

\section{Discussion}
\label{sec:discussions}

\subsection{Limitations of the Emulation Techniques}
\label{sec:discuss-limit}

Although our approach is able to discover vulnerabilities in embedded web interfaces 
that run inside an emulated environment, setting up such environments is not always
straightforward. We discuss several limitations we encountered and 
outline how they could be handled in the future.
In fact, many of these limitations are the results of the failures 
analyzed previously in Section~\ref{sec:results-failures}. 

\subsubsection{Forced Emulation}
\label{sec:discuss-forcedemulation}

Even though most of the firmware instances in our dataset are for Linux-based devices, they are 
quite heterogeneous and their actual binaries vary. Examples include 
\texttt{init} programs that have different set of command parameters or strictly 
requiring to run as PID 0, which is not the case in a chrooted environment.
Ideally, there should be a simple and uniform way to start the firmware, but this is not the case in practice 
as devices are very heterogeneous. In addition to this, unless we have access to the 
bootloader of each individual device, there is no 
reliable way to reproduce the boot sequence. Obtaining and 
reverse-engineering the bootloaders themselves is not trivial. This usually 
requires access to the device, use of physical memory dumping techniques, and 
manual reverse-engineering, which is outside the scope of this paper. 
We emulate firmware images by forcefully invoking its default
initialization scripts, (e.g., \texttt{/etc/init}, \texttt{/etc/rc}),
however, sometimes, these scripts do not exist or fail to execute correctly 
leading to an incomplete  system configuration.
For instance, it may fail to mount the \texttt{/etc\_ro} partition at
the \texttt{/etc} mount point, and then, the web server is missing some
required files (e.g., \texttt{/etc/passwd}).

\subsubsection{Emulated Web Server Environment}
\label{sec:discuss-emulatedwebserverenv}

Even if the basic emulation was successful, other problems
with the emulated web server environment are common. 
For example, an emulated web interface return for 
many requests the HTTP response codes 
\texttt{500 Internal Server Error} or \texttt{404 Not Found}. 
Manual inspection of the cases when code \texttt{500} is returned suggests that 
some scripts or binaries are either missing from the root filesystem or 
do not have proper permissions. 
Code \texttt{404} was often returned due to the wrong  
web server configuration file being loaded, which lead to  
the document root pointing at a wrong directory. To 
overcome this, we try to emulate the web interface of a firmware using 
all combinations of the configuration files and document roots 
we find in this firmware.

\subsubsection{Imperfect Emulation}
\label{sec:discuss-imperfectemu}

The ability to emulate embedded software in QEMU is incredibly
valuable, but comes at a price. One big drawback is that some
very basic peripheral devices are missing in the emulated
environments. 
A very common emulation failure is related to the lack of non
volatile memories (e.g., NVRAM)~\cite{devttyS0-nvram-emulate, shadowfile-nvram-emulate}. 
Such memories are used by
embedded devices to store boot and configuration information. 
Several approaches to overcome such limitations exist. One is to
have an universal or on-the-fly NVRAM emulator plugged into QEMU, 
for example instrumented at kernel-level or 
implemented using \texttt{Avatar}~\cite{zaddach:ndss14}.  Another
approach is to intercept calls to the commonly used \texttt{libnvram}
functions (such as \texttt{nvram\_get} and \texttt{nvram\_set}) and
return fake data~\cite{devttyS0-nvram-emulate, shadowfile-nvram-emulate}. 
While these tools are easy to compile and use, it is not trivial to 
automatically generate meaningful application data without producing 
false alerts or breaking the emulation.
We plan to integrate these techniques in our future versions.

\subsection{Outdated Firmware Versions}
\label{sec:discus-outdated}

One concern about our approach could be that the firmware files in 
our experiments were not necessarily the latest available versions. 
This in turn could imply that the vulnerabilities we automatically 
discovered are not necessarily applicable to the latest versions 
of the affected firmware images. Although such a concern is legitimate, 
in practice there are several caveats to this concern that in our 
view still make our methodology and findings valuable. 

    First, it is important to know and understand how many embedded devices 
that are vulnerable or have outdated firmware will update their firmware 
in such a case. 
On the one hand, many embedded devices are SOHO devices which means that the 
users decide \textit{if and when} they will upgrade their firmware version. 
On the other hand, researchers showed that even simple improvements, 
such as changing the default credentials of the embedded devices, are 
not always applied by the users during long period of times~\cite{ghena2014green}. 
For example, it was found that 96\% of accessible devices 
having factory default root passwords still remain vulnerable after 
a period of 4 months~\cite{cui-acsac2010-QuantAnalysInsecEmbDev}. 
On the other hand, a firmware download and update is a more complex task 
than a change of the default credentials. 
Therefore, unless the devices are connected to the Internet and have 
a firmware auto-update functionality that is effectively \emph{enabled}, it 
is reasonable to expect that in practice the firmware updates are applied far 
less than expected/desired, or are applied at best as often and as fast as 
the credentials are updated. 

    Second, even though the embedded devices should keep their firmware 
updated, this in not always feasible, e.g., for field-deployed devices. 
Such devices often cannot be remotely updated and require the physical 
access of an operator in the field to do so. However, even in such cases 
the upgrade of the firmware is not always straightforward. 
Cerrudo~\cite{cerrudo2014hacking} showed that in some cases embedded 
devices could be buried in the roadway, making firmware updates that 
require physical access very challenging, if not impossible. 

    Third, even the latest firmware releases could still contain 
the very same vulnerabilities as the older versions~\cite{cui-ndss2013-FirmwareModAttacks-old}. 
Therefore, vulnerabilities discovered in older firmware versions can 
prove extremely useful as direct input or mutation template for testing 
the latest firmware versions. 

In summary, we believe that a security study performed only on the 
latest firmware releases could provide important details for securing 
embedded devices (e.g., critical vulnerability discovery, patching 0-days). 
At the same time, however, such a study would not be completely accurate 
as many existing devices run outdated firmware versions. 
Ultimately, the goal of this work is not to find (all) the vulnerabilities 
in (all) the latest firmware versions. The main goal is to provide 
a methodology and insights that can be applied on any firmware version 
in order to automatically discover vulnerabilities in embedded firmware, 
and in particular in embedded web interfaces.

\subsection{Manual Interventions}
\label{sec:discus-manual}

Our framework is designed to be as automated as possible.  However,
manual interventions are sometimes necessary or even desirable.
    First, for each newly encountered web server type we need to write a
tool, which will then automatically detect, parse, and launch instances
of this particular web server. Automation of such a step could be
improved, for example, using \texttt{ConfigRE}~\cite{wang2008towards}.
    Second, manual inspection of the results and of the affected software
allows to confirm vulnerabilities and sometimes leads to finding new ones. This
is part of the power of our methodology, i.e., pointing the finger on likely
vulnerable software. In our experience this last phase was very
productive as there were only a few false positives left after the dynamic
analysis phase.

\subsection{Ethical Aspects}
\label{sec:discus-ethical}

In our study we are particularly careful to work within legal and ethical
boundaries. 
    First, we strictly follow the \emph{responsible disclosure} policy. 
To this end, we try our best to notify vendors, CERTs and 
Vulnerability Contribution Programs (VCP) for vulnerabilities 
we discover during our experiments. We also try to assist vendors in 
reproducing these issues. 
    Second, as previously mentioned, our framework does not operate on 
live embedded devices, rather on their emulations. 
This avoids both accessing devices we do not own and 
breaching the terms of use.  Also, there is no risk to interfere
unintentionally with devices which are not under our control or to
``brick'' an actual device. 
In limited cases when confirmation of an
issue requires a physical device, we do perform such validations on
devices under our control and in an isolated test environment.

\section{Related Work}
\label{sec:related-work}

Analysis of embedded devices is not a new idea.
Costin et al.~\cite{costin2014large} preformed a large scale analysis but they 
did it only through a simple static analysis.
Bojinov et al.~\cite{elie-bh2009-EmbedInterfMassInsec} studied the security of
embedded management interfaces but performed the analysis manually on
only 21 devices. Similar studies were recently preformed on popular
SOHO devices~\cite{seceval-trendnet, hpfortifyiot2014} each performing
manual analysis on about 10 devices and uncovering flaws in them. In
contrast to these, we show that by automating the analysis we can scale
to testing hundreds of devices and find thousands of vulnerabilities.

In addition, several projects scanned the Internet, or parts of it, to discover
vulnerabilities in embedded systems~\cite{cui-raid2009-BraveNewWorld,
cui-acsac2010-QuantAnalysInsecEmbDev, upnp_scan, internetcensus2012,
shodan, psqs2012}. In most cases these approaches lead to discovery
of devices with known vulnerabilities such as default passwords or
keys, and in several notable cases helped the discovery of new flaws~\cite{psqs2012}. However,
such approaches raise serious ethical problems
and in general only allow to find devices that are vulnerable to known
(manually found) bugs.

Web static analysis is a very active field of research, Huang et
al.~\cite{Huang:2004:SWA:988672.988679} were the first to statically
search for web vulnerabilities in the context of PHP
applications. 
Balzarotti et al.~\cite{balzarotti2008saner} showed that even if the
developer performs certain sanitization on input data, XSS attacks are
still possible due to the deficiencies in the sanitization
routines.  
Pixy~\cite{jovanovic2006pixy} proposed a technique based on data flow
analysis for detecting XSS, SQL or command injections.
RIPS~\cite{Dahse:rips:ndss14}, on the other hand, is a static code
analysis tool that detects multiple types of injection
vulnerabilities.  
While in this work we can, in principle use any of those
detection mechanisms we only used RIPS which has low false positives
and is still openly available.

There are several recent works that rely on emulation in order to discover 
or verify vulnerabilities in embedded systems. 
    Zaddach et al.~\cite{zaddach:ndss14} proposed \texttt{Avatar}, which is a dynamic 
analysis framework for firmware security testing of embedded devices. 
\texttt{Avatar} executes embedded code inside a QEMU emulator, while the 
I/O requests to the peripherals of the embedded system are forwarded to
the real device attached to the framework. 
Kammerstetter et al.~\cite{prospect} targeted Linux-based embedded
systems that are emulated with a custom kernel which forwards
\texttt{ioctl} requests to the embedded device that runs the normal
kernel.  Li et al.~\cite{li2010femu} proposed a hybrid
firmware/hardware emulation framework to achieve confident SoC
verification. Authors used a transplanted QEMU at BIOS level to
directly emulate devices upon hardware.  Unfortunately, those
approaches are not possible to use without having access to a
physical device, which does not scale as our approach does. 

Meanwhile, Shoshitaishvili et al.~\cite{firmalice2015ndss} presented
\texttt{Firmalice}, a static binary
analysis framework to support the analysis of firmware files for
embedded devices. It was demonstrated to detect three \emph{known
  backdoors} in real devices, but it requires manual annotations and
is therefore not possible to use in a large scale analysis.

Fong and Okun~\cite{fong2007web} took a closer look at web application 
scanners, and their functions and definitions. 
Bau et al.~\cite{elie-sp2010-AutoBlackBoxWebTest} conducted an evaluation 
of the state of the art of tools for automated ``black box'' web application 
vulnerability testing. While results have shown the promise and effectiveness 
of such tools, they also uncovered many limitations of existing tools. 
Similarly, Doup\'e et al.~\cite{doupe2010johnny} performed an evaluation 
of eleven ``black box'' web pen-testing tools, both open-source and commercial. 
Authors found that crawling ability is as important and challenging as 
vulnerability detection techniques and many classes of vulnerabilities 
are completely overlooked. They conclude that more research is required 
to improve the tools. 
Holm et al.~\cite{holm2011quantitative} performed a quantitative evaluation 
of vulnerability scanning. Authors showed that automated scanning is unable 
to accurately identify all vulnerabilities. They also showed that scans of 
Linux-based hosts are less accurate than the scans of Windows-based ones. 
Doup\'e et al.~\cite{doupe2012enemy} proposed improvements to the ``black box'' 
vulnerability testing tools. Authors observed the web application state 
\emph{from the outside}. This allowed them to extend their testing coverage. 
Then they drove the ``black box'' web application vulnerability scanner. 
They implemented the technique in a crawler linked to a fuzzing 
component of an open-source web pen-testing tool. 
Such improvements to analysis tools will benefit our framework as we can 
integrate them in our analysis phase. 

Finally, Gourdin et al.~\cite{elie-usenix2011-SecEmbWebInterf}
addressed the challenges of building secure embedded web interfaces by
proposing \texttt{WebDroid}, which was the first framework
specifically dedicated to this purpose.

\section{Conclusion and Future Work}
\label{sec:conclusion}

In this work, we presented a new methodology to perform large scale
security analysis of web interfaces within embedded devices.  
For this, we designed the first framework that achieves \emph{scalable and automated 
dynamic analysis of firmwares}, and that was precisely developed to 
discover vulnerabilities in embedded devices using the software-only approach. 
Our framework leverages off-the-shelf static and 
dynamic analysis tools. Because of the limitations in static analysis tools, we
created a mechanism for automatic emulation of firmware images.  While
perfectly emulating unknown hardware will probably remain an open
problem, we were able to emulate systems well enough to test the web
interfaces of \CountFirmwareEmulatedWebServerStarted{} firmware
images. Our framework found serious vulnerabilities in at least 
\printpercent{\CountFirmwareDynVulnTotal}{\CountFirmwareEmulatedWebServerStarted} of the web interfaces we were able to emulate,  
including \CountDynVulnsTotalHigh{} \emph{high impact} vulnerabilities 
found and verified by dynamic analysis. 
When counting static analysis,
\CountVulnsConclusion{} issues were found in 
\countfirmwarestotal{} firmware images, affecting nearly a quarter of vendors in our dataset. 
These results show that some embedded systems manufacturers need to
start considering security in their software life-cycle, e.g.,
using off-the-shelf security scanners as part of their product quality assurance.

Our work motivates the need for additional research in several
areas. First, there are probably ways to improving emulation quality
of unknown hardware. Second, automatically synthesizing web exploits
would make vulnerability confirmation easier. Finally, responsibly
disclosing vulnerabilities is time consuming and difficult (and in our
experience is worse with vendors of SOHO devices). It becomes an open
challenge when it needs to be performed at a large scale.

We plan to continue collecting new data and extend our analysis to all
the firmware images we can access in the future.  Further we want to
extend our system with more sophisticated dynamic analysis techniques
that allow a more in-depth study of vulnerabilities within each
firmware image.

\balance

\bibliographystyle{abbrv}
\small{
\bibliography{./firmware-genomics}
}

\end{document}